\documentclass[twocolumn]{aastex701} 

\usepackage[hyphens]{url}
\usepackage{multirow}
\usepackage{caption}
\begin{document}

\title{A Large Sample of JWST/NIRSpec Brown Dwarfs: New Distant Discoveries}

\correspondingauthor{Shu Wang}
\email{shuwang@nao.cas.cn}

\author[0009-0000-7976-7383]{Zhijun Tu}
\affiliation{CAS Key Laboratory of Optical Astronomy, National Astronomical Observatories, Chinese Academy of Sciences, Beijing 100101, People's Republic of China}
\email{zjtu@bao.ac.cn}

\author[0000-0003-4489-9794]{Shu Wang}
\affiliation{CAS Key Laboratory of Optical Astronomy, National Astronomical Observatories, Chinese Academy of Sciences, Beijing 100101, People's Republic of China}
\affiliation{School of Astronomy and Space Sciences, University of Chinese Academy of Sciences, Beijing 100049, People's Republic of China}
\email{shuwang@nao.cas.cn}

\author[0000-0001-7084-0484]{Xiaodian Chen}
\affiliation{CAS Key Laboratory of Optical Astronomy, National Astronomical Observatories, Chinese Academy of Sciences, Beijing 100101, People's Republic of China}
\affiliation{School of Astronomy and Space Sciences, University of Chinese Academy of Sciences, Beijing 100049, People's Republic of China}
\affiliation{Institute for Frontiers in Astronomy and Astrophysics, Beijing Normal University, Beijing 102206, People's Republic of China}
\email{chenxiaodian@nao.cas.cn}

\author{Jifeng Liu}
\affiliation{CAS Key Laboratory of Optical Astronomy, National Astronomical Observatories, Chinese Academy of Sciences, Beijing 100101, People's Republic of China}
\affiliation{School of Astronomy and Space Sciences, University of Chinese Academy of Sciences, Beijing 100049, People's Republic of China}
\affiliation{Institute for Frontiers in Astronomy and Astrophysics, Beijing Normal University, Beijing 102206, People's Republic of China}
\affiliation{New Cornerstone Science Laboratory, National Astronomical Observatories, Chinese Academy of Sciences, Beijing 100012, People's Republic of China}
\email{jfliu@nao.cas.cn}

\begin{abstract}
Brown dwarfs are essential probes of stellar and planetary formation, yet their low luminosities pose challenges for detection at large Galactic distances. The James Webb Space Telescope (JWST), with its unprecedented near-infrared sensitivity, enables the discovery and characterization of distant substellar objects, including those in the Milky Way's thick disk and halo. We conducted a systematic search using over 40,000 publicly available JWST/NIRSpec PRISM/CLEAR spectra and identified 68 brown dwarfs through spectral template matching and visual inspection. Among them, 12 are newly identified candidates, including 8 T dwarfs and 4 M/L dwarfs, most at distances exceeding 1 kpc. Remarkably, two sources---JWST J001418.22$-$302223.2 and JWST J033240.07$-$274907.8---are found at distances greater than 5\,kpc, making them the most distant brown dwarfs within the Milky Way. Spectral fits were performed using a nested sampling Monte Carlo algorithm with three model grids: Sonora Elf Owl, LOWZ, and SAND. The analysis reveals that cloud-free models are unable to reproduce L/T transition spectra, whereas the SAND model provides a more accurate representation of cloud effects in metal-poor environments. With the newly identified distant brown dwarfs, we also investigated the vertical metallicity gradient of brown dwarfs. Overall, the metallicities do not show an evident trend with Galactic height $|Z|$, due to the limited sample size and the uncertainties in metallicity measurements.

\end{abstract}


\section{Introduction} \label{sec:intro}

Brown dwarfs, with masses lower than the threshold for sustained core hydrogen fusion ($\lesssim$0.075\,$M_\odot$, \citealt{Burrows2001RvMP}), represent a key population for understanding the low-mass end of star formation, atmospheric physics under extreme conditions, and the boundary between stellar and planetary regimes. The large-scale infrared surveys such as 2MASS, UKIDSS, and WISE have enabled the discovery of hundreds of L, T, and even the coldest class of Y dwarfs in the Solar neighborhood ($\lesssim20\,\text{pc}$), greatly expanding the known brown dwarf population and spectral diversity over the past two decades \citep[e.g.,][]{1999ApJ...519..802K,2000AJ....120..447K,Cushing2011ApJ,2013MNRAS.433..457B}. However, their intrinsically low temperatures ($T_\mathrm{eff}<2400\,\text{K}$) and luminosities make them challenging to detect, especially at large distances ($>100\,\text{pc}$) or in crowded, high-density fields.

The \textit{James Webb Space Telescope} (JWST), with its unprecedented sensitivity and wavelength coverage in the near- to mid-infrared, provides a transformative opportunity for the study of brown dwarfs \citep[e.g.,][]{Beiler2023ApJ,2023ApJ...959...86L,2024ApJ...976...82T}. In particular, the NIRSpec instrument offers numerous high-quality, low/medium-resolution spectroscopy in the 0.6--5.3\,$\mu$m range, enabling detailed spectral characterization of cool atmospheres, including key molecular absorption features such as H$_2$O, CH$_4$, CO$_2$, and CO \citep[e.g.,][]{Beiler2024ApJ,2024AJ....167..237L}. These capabilities allow not only for improved knowledge of known brown dwarfs but also for the discovery of previously unidentified brown dwarfs in archival datasets \citep[e.g.,][]{Burgasser2024ApJ,2025MNRAS.tmpL..66L}. Expanding the census of brown dwarfs will provide critical constraints on their population statistics and formation mechanisms, and will aid in refining theoretical models of their internal structure and atmospheric chemistry.

While dedicated observations targeting nearby brown dwarfs have significantly advanced our understanding of substellar atmospheres, JWST's extensive extragalactic programs have unexpectedly emerged as a valuable resource for brown dwarf science. Previous studies based on deep imaging from the \textit{Hubble Space Telescope} (HST) and JWST have shown that brown dwarfs can exhibit broadband colors similar to those of high-redshift galaxies and compact red sources—often referred to as ``little red dots'' (LRDs)—in deep extragalactic fields \citep[e.g.,][]{Langeroodi2023ApJ,2024ApJ...964...66H,2025ApJ...980..230T}. This color degeneracy has led to instances where brown dwarfs were misclassified or initially identified as extragalactic objects, particularly in galaxy evolution surveys, which are a major focus of JWST observations. With the advent of JWST, a new generation of deep spectroscopic surveys—such as JADES \citep{JADES}, CEERS \citep{CEERS}, and UNCOVER \citep{UNCOVER}—has dramatically expanded the coverage and depth of near-infrared data across key extragalactic fields. Although these programs were primarily designed to probe early galaxy populations and cosmic reionization, they also provide valuable spectroscopic information on foreground Galactic objects, including brown dwarfs. In addition, early results from the Euclid Quick Release 1 (Q1) data have demonstrated the ability to detect and characterize ultracool dwarfs within the Euclid Deep Fields survey, further highlighting the potential for substellar discoveries in wide-area extragalactic surveys \citep{Euclid1,Euclid2,Euclid3}.

Given the large volume of public JWST/NIRSpec spectra now available, it is highly likely that many Galactic brown dwarfs remain hidden or unrecognized in these datasets \citep{2016AJ....151...92R}. This motivates a systematic search for previously unidentified brown dwarfs within archival NIRSpec observations. Identifying such distant brown dwarfs in deep extragalactic fields not only expands the empirical census of substellar objects, but also provides opportunities to study brown dwarfs in the Galactic thick disk and halo populations. Such brown dwarfs are crucial for probing the low-mass end of the stellar initial mass function in old Galactic components and can provide dynamical and chemical constraints on the structure and evolution of the Milky Way \citep[e.g.,][]{2003ApJ...592.1186B,2004ApJ...614L..73B,2018MNRAS.479.1383Z}.

Once stellar metallicities at different distances are determined, the metallicity-vertical height ($|Z|$) relation can be constructed. Previous studies based on stellar populations consistently report a negative vertical metallicity gradient, with lower metallicities at larger $|Z|$, reflecting older, dynamically heated populations such as the thick disk and halo \citep{1983MNRAS.202.1025G, 2000AJ....119.2843C,2011AJ....142..184C,2012ApJ...761..160S,2021ApJ...923..145W,2025ApJ...983...51L}. This gradient is widely interpreted as evidence for the inside-out and upside-down formation of the Galactic disk \citep{2013ApJ...773...43B,2014A&A...572A..92M}. In contrast, analogous studies for brown dwarfs remain scarce due to their intrinsic faintness and the lack of known objects, particularly at kiloparsec distances, which have only recently become accessible through new discoveries. In this work, by identifying a substantial sample of distant brown dwarfs, we aim to provide the first preliminary investigation of the metallicity-$|Z|$ relation in the substellar regime.

This paper is organized as follows. Section \ref{sec:data} describes the datasets used in this study, while Section \ref{sec:method} outlines the methods used to identify brown dwarfs within these data. Section \ref{sec:spec_analysis} details the modeling approaches and analysis techniques adopted to derive the physical parameters of the brown dwarfs. In Section \ref{sec:results}, we present the derived physical parameters, spectral classifications, and distance estimates. Section \ref{sec:discuss} focuses on the investigation of the vertical metallicity gradient in our brown dwarf sample. Finally, Section \ref{sec:conclusion} summarizes our main findings and highlights the implications of this work.

\begin{deluxetable*}{llrccccc}
\setlength{\tabcolsep}{0.1cm}
\tabletypesize{\footnotesize}
\tablecaption{Observational Information of Brown Dwarf Candidates\label{tab:sample}}
\tablehead{\colhead{Object Name} & \colhead{$l$} & \colhead{$b$} & \colhead{PID} & \colhead{PI} &\colhead{Source Number} & \colhead{Project} & \colhead{Exposure Time} \\ 
\colhead{} & \colhead{(deg)} & \colhead{(deg)} & \colhead{} & \colhead{} & \colhead{} & \colhead{} & \colhead{(s)} } 

\startdata
JWST J001418.22$-$302223.2 & 9.004385 & $-$81.241781 & 2561 & I. Labbe & 41160 & UNCOVER & 16019 \\
JWST J021651.35$-$050623.6 & 169.465596 & $-$60.018034 & 4233 &A. G. de Graaff& 170428 & RUBIES/PRIMER & 2889 \\
JWST J021701.94$-$050621.7 & 169.531830 & $-$59.989092 & 4233 &A. G. de Graaff& 170824 & RUBIES/PRIMER & 963 \\
JWST J021715.61$-$050857.7 & 169.674308 & $-$59.985173 & 4233 &A. G. de Graaff& 140125 & RUBIES/PRIMER & 2889 \\
JWST J031347.94$-$671235.0 & 284.511189 & $-$44.485687 & 1747 &G. Roberts-Borsani& 676 & SuperBoRG & 540 \\
JWST J033225.12$-$274636.1 & 223.516653 & $-$54.443205 & 1180 &D. J. Eisenstein& 39095 & JADES/GOODS & 1255 \\
JWST J033240.07$-$274907.8 & 223.602385 & $-$54.395010 & 1180 &D. J. Eisenstein& 197182 & JADES/GOODS & 1036 \\
JWST J065810.66$-$555659.5 & 266.012961 & $-$21.293910 & 4598 &M. Bradac& 7103870 & * & 7090 \\
JWST J091049.86$-$041320.8 & 234.612343 & 28.267592 & 2028 &F. Wang& 14911 & * & 2218 \\
JWST J141916.32$+$525258.8 & 96.527636 & 59.508137 & 4106 &E. Nelson& 87141 & CEERS & 4027 \\
JWST J141919.36$+$525006.9 & 96.454438 & 59.539153 & 1345 &S. L. Finkelstein& 1558 & CEERS & 3107 \\
JWST J220249.96$+$185059.6 & 76.658171 & $-$28.496439 & 1747 &G. Roberts-Borsani& 802 & SuperBoRG & 540 \\
\enddata
\tablecomments{The Object Name is formatted as ``hhmmss.ss+ddmmss.s''. $l$ and $b$ are the Galactic longitude and latitude, respectively. PID refers to the JWST program ID from which the source was obtained, and Source Number indicates the source identifier within the corresponding PID. Project Name denotes the specific observing program or project associated with the source’s location; an asterisk (*) indicates that no specific project name is available for that position. Exposure Time represents the total exposure time of the obtained spectrum.}
\end{deluxetable*}

\section{Data} \label{sec:data}

Among the available JWST/NIRSpec spectroscopic modes, the NIRSpec PRISM/CLEAR configuration currently provides the largest volume of publicly released data. We therefore focus our analysis on this mode. The PRISM/CLEAR setting offers low-resolution spectroscopy ($R \sim 100$) over the entire near-infrared wavelength range from 0.6 to 5.3\,$\mu$m, making it well suited for identifying the broad molecular absorption features that characterize brown dwarf spectra, particularly around $\sim1$--2.4 and $\sim4\,\mu$m.

To construct our sample, we retrieved all publicly available stage 3 NIRSpec PRISM/CLEAR spectra from the Mikulski Archive for Space Telescopes (MAST)\footnote{\url{https://mast.stsci.edu/search/ui/\#/jwst}}, as of early April 2025. Stage 3 products are fully calibrated, combined spectra that are ready for scientific analysis. In total, we obtained 41283 individual spectra from a wide range of programs, including deep extragalactic surveys, calibration programs, and numerous dedicated brown dwarf observations.

All of these data downloaded from MAST contain two columns that provide the source information: \textit{targdesc}, which describes the specific subclass of the source, and \textit{targcat}, which describes the broader category. For example, a known brown dwarf may be labeled in \textit{targdesc} as ``brown dwarf'', or more specifically as ``T dwarf'' or ``Y dwarf'', but in \textit{targcat} all such known brown dwarfs are uniformly labeled as ``Star''. We examined all 41283 sources and found that only a very small fraction ($\sim$2\%) are marked as known objects. Among these, 124 are labeled as ``Galaxy'', 11 as ``ISM'', 175 as ``Solar System'', 164 as ``Star'', and 402 as ``Calibration''. 
Notably, the ``Calibration'' category also includes many A-type stars. Therefore, the identification of these unknown objects in the JWST sample is both meaningful and necessary.

\section{Methods and Sample} \label{sec:method}

To systematically select brown dwarfs and brown dwarf candidates within the dataset, we performed spectral model fitting using the brown dwarf atmospheric models. For each observed spectrum, we computed the reduced chi-square ($\chi_r^2$) relative to the model grid, identifying the best-fit model as the one that minimizes $\chi_r^2$. The $\chi_r^2$ is defined as:
\begin{equation}
    \chi_r^2 = \frac{\chi^2}{N - p} = \sum_i \frac{(F_i - \frac{R^2}{D^2} M_i)^2}{\sigma_i^2 (N - p)},
\end{equation}
where $F_i$ and $\sigma_i$ are the observed flux and its uncertainty at each wavelength point $i$, and $M_i$ is the corresponding model flux. Here, $N$ is the number of spectral data points, and $p$ is the number of free model parameters. The scaling factor $R^2/D^2$, which accounts for the ratio of the source radius to its distance, is derived analytically by minimizing $\chi^2$:
\begin{equation}
    \frac{R^2}{D^2} = \frac{\sum_i \left(F_i M_i / \sigma_i^2 \right)}{\sum_i \left(M_i^2 / \sigma_i^2 \right)}.
\end{equation}

We adopted the Sonora-Bobcat model grids \citep{Marley2021ApJ}, which provide synthetic spectra for a wide range of effective temperatures, surface gravities, and metallicities representative of L, T, and Y-type brown dwarfs. The grid spans:
\begin{itemize}
    \item $T_\mathrm{eff}$ from 200 to 2400\,K, with a sampling interval of 25\,K below 600\,K, 50\,K between 600 and 1000\,K, and 100\,K above 1000\,K;
    \item surface gravity ranges from $\log g = 3.0$ to 5.5 (in cgs units) in steps of 0.25 dex;
    \item metallicities of [M/H] = $-0.5$, 0.0, and $+0.5$ are available;
    \item A subset of the models also explores variations in the carbon-to-oxygen (C/O) ratio, with values of C/O = 0.23 and 0.69.
\end{itemize}

We resample and convolve the model spectra to match the spectral resolution of NIRSpec PRISM/CLEAR, and choose the one that minimizes the $\chi_r^2$ as the best-fit model of the observed spectrum. 

We first visually inspected all 41283 spectra by comparing them with their best-fit models from the Sonora-Bobcat grid, focusing on diagnostic regions for cool brown dwarfs such as the 1.0, 1.25, 1.6, and 2.1\,$\mu$m flux peaks shaped by prominent H$_2$O and CH$_4$ absorption bands, as well as the overall spectral fit. From this procedure, we retained as brown dwarf candidates those sources whose best-fit effective temperatures did not converge to the upper boundary of the Sonora-Bobcat grid ($T_\mathrm{eff} = 2400$\,K). Subsequently, we cross-checked all sources against the \textit{targdesc} column in MAST and re-incorporated those labeled as brown dwarfs that had not been included in our initial visual selection (e.g., sources with $T_\mathrm{eff} \geq 2400$\,K).

Through a careful inspection and the above procedure, we conducted a sample of 68 brown dwarfs, including 12 newly identified candidates. The detailed observational properties of these 12 new candidates are provided in Table~\ref{tab:sample}. We note that all of these candidates were selected from extragalactic deep field surveys and are located at Galactic latitudes $|b| > 20^\circ$. The observational information for the remaining 56 identified brown dwarfs is presented in Table~\ref{tab:oldsample}.

\startlongtable
\begin{deluxetable*}{clrccc}
\setlength{\tabcolsep}{0.5cm}
\tablecaption{Observational Information of the Identified Brown Dwarfs\label{tab:oldsample}}

\tablehead{\colhead{Object Name} & \colhead{R.A.} & \colhead{Decl.} & \colhead{PID} & \colhead{PI} & \colhead{Reference$^a$} \\ 
\colhead{} & \colhead{(deg)} & \colhead{(deg)} & \colhead{} & \colhead{} & \colhead{} } 
\startdata
2MASS J00470038$+$6803543 & 11.758509 & 68.063718 & 3486 & J. Vos &  \\
WISE J024714.52$+$372523.5 & 41.810844 & 37.422775 & 2302 & M. C. Cushing & (1) (2) (11)\\
WISEPA J031325.96$+$780744.2 & 48.360019 & 78.129222 & 2302 & M. C. Cushing & (1) (2) (11)\\
JADES-GS-BD-9 & 53.161143 & $-$27.809174 & 1180 & D. J. Eisenstein & (9) \\
IC 348 MM 51 & 56.187319 & 32.143752 & 1229 & C. Alves de Oliveira & (3) \\
2MASS J034807.72$-$602227.0 & 57.028837 & $-$60.378992 & 1189 & T. L. Roellig & (2) (11)\\
2MASS J03552337$+$1133437 & 58.848881 & 11.557916 & 3486 & J. Vos &  \\
WISE J035934.06$-$540154.6 & 59.891347 & $-$54.034355 & 2302 & M. C. Cushing & (1) (2) (4) (11)\\
WISE J043052.92$+$463331.6 & 67.725061 & 46.560044 & 2302 & M. C. Cushing & (1) (2) (11)\\
ONC J083.77378$-$05.37957 & 83.773765 & $-$5.379561 & 2770 & M. J. McCaughrean & (10) \\
ONC J083.78378$-$05.37705 & 83.783807 & $-$5.377038 & 2770 & M. J. McCaughrean & (10) \\
ONC J083.78472$-$05.38936 & 83.784765 & $-$5.389361 & 2770 & M. J. McCaughrean & (10) \\
ONC J083.78535$-$05.40039 & 83.785337 & $-$5.400403 & 2770 & M. J. McCaughrean & (10) \\
ONC J083.78972$-$05.33420 & 83.789722 & $-$5.334233 & 1228 & C. Alves de Oliveira & (5) \\
ONC J083.80006$-$05.38467 & 83.800039 & $-$5.384671 & 2770 & M. J. McCaughrean & (10) \\
ONC J083.80445$-$05.34607 & 83.804455 & $-$5.346102 & 1228 & C. Alves de Oliveira & (5) \\
ONC J083.81412$-$05.34651 & 83.814123 & $-$5.346519 & 1228 & C. Alves de Oliveira & (5) \\
COUP 714 & 83.814506 & $-$5.328756 & 1228 & C. Alves de Oliveira & (5) \\
WISE J053516.80$-$750024.9 & 83.818507 & $-$75.00667 & 2302 & M. C. Cushing & (1) (2) (11)\\
ONC J083.81892$-$05.44293 & 83.818931 & $-$5.442934 & 2770 & M. J. McCaughrean & (10) \\
ONC J083.82316$-$05.43600 & 83.823165 & $-$5.436 & 2770 & M. J. McCaughrean & (10) \\
ONC J083.82323$-$05.37439 & 83.823222 & $-$5.37439 & 1228 & C. Alves de Oliveira & (5) \\
ONC J083.82387$-$05.36243 & 83.823868 & $-$5.362454 & 1228 & C. Alves de Oliveira & (5) \\
ONC J083.82400$-$05.32450 & 83.824004 & $-$5.324522 & 1228 & C. Alves de Oliveira & (5) \\
ONC J083.82520$-$05.36140 & 83.825203 & $-$5.361437 & 1228 & C. Alves de Oliveira & (5) \\
ONC J083.82622$-$05.36372 & 83.826224 & $-$5.363731 & 1228 & C. Alves de Oliveira & (5) \\
ONC J083.83180$-$05.40876 & 83.831815 & $-$5.408753 & 2770 & M. J. McCaughrean & (10) \\
ONC J083.83189$-$05.41171 & 83.831904 & $-$5.411724 & 2770 & M. J. McCaughrean & (10) \\
ONC J083.83198$-$05.37616 & 83.831989 & $-$5.37618 & 1228 & C. Alves de Oliveira & (5) \\
ONC J083.83626$-$05.41327 & 83.836251 & $-$5.413269 & 2770 & M. J. McCaughrean & (10) \\
2MASS J06420559$+$4101599 & 100.523281 & 41.030589 & 3486 & J. Vos &  \\
2MASS J06462756$+$7935045 & 101.6107 & 79.580365 & 3670 & B. Burningham &  \\
WISE J073444.02$-$715744.0 & 113.676648 & $-$71.962471 & 2302 & M. C. Cushing & (1) (11)\\
WISE J082507.35$+$280548.5 & 126.280454 & 28.095954 & 2302 & M. C. Cushing & (1) (2) (11)\\
WISEA J085510.74$-$071442.5 & 133.765779 & $-$7.242687 & 1230 & C. Alves de Oliveira & (6) (11) \\
ULAS J102940.52$+$093514.6 & 157.417185 & 9.586871 & 2302 & M. C. Cushing & (1) (2) \\
CWISEP J104756.81$+$545741.6 & 161.98458 & 54.961238 & 2302 & M. C. Cushing & (1) (2) (11)\\
WISE J120604.38$+$840110.6 & 181.499192 & 84.018605 & 2302 & M. C. Cushing & (1) (2) (11)\\
Ross 458C & 195.171147 & 12.353974 & 1277 & P.-O. Lagage &  \\
SDSSp J134646.45$-$003150.4 & 206.690035 & $-$0.531438 & 2302 & M. C. Cushing & (1) \\
WISEPC J140518.40$+$553421.4 & 211.312046 & 55.573408 & 2302 & M. C. Cushing & (1) (2) (11)\\
o005\_s41280 & 214.828485 & 52.81083 & 4233 & A. G. de Graaff & (7) \\
o006\_s00089 & 214.910295 & 52.860069 & 4233 & A. G. de Graaff & (7) \\
o006\_s35616 & 214.938851 & 52.873855 & 4233 & A. G. de Graaff & (7) \\
CWISEP J144606.62$-$231717.8 & 221.525515 & $-$23.290376 & 2302 & M. C. Cushing & (1) (2) (11)\\
WISE J150115.92$-$400418.4 & 225.318306 & $-$40.073014 & 2302 & M. C. Cushing & (1) (2) (11)\\
WISEPA J154151.66$-$225025.2 & 235.461629 & $-$22.840795 & 2302 & M. C. Cushing & (1) (2) (11)\\
SDSS J162414.37$+$002915.6 & 246.05738 & 0.487635 & 2302 & M. C. Cushing & (1) (2) \\
2MASS J17410280$-$4642218 & 265.261431 & $-$46.708506 & 3486 & J. Vos &  \\
WISEPA J182831.08$+$265037.8 & 277.133396 & 26.844449 & 1189 & T. L. Roellig & (8) (11)\\
WISEPA J195905.66$-$333833.7 & 299.773538 & $-$33.64342 & 2302 & M. C. Cushing & (1) (2) \\
WISEPC J205628.90$+$145953.3 & 314.123306 & 14.999958 & 2302 & M. C. Cushing & (1) (2) \\
WISE J210200.15$-$442919.5 & 315.500698 & $-$44.490045 & 2302 & M. C. Cushing & (1) (2) (11)\\
WISEA J215949.54$-$480855.2 & 329.957966 & $-$48.152953 & 2302 & M. C. Cushing & (1) (2) (11)\\
WISE J220905.73$+$271143.9 & 332.278519 & 27.190841 & 2302 & M. C. Cushing & (1) (2) \\
WISEA J235402.79$+$024014.1 & 358.513404 & 2.669219 & 2302 & M. C. Cushing & (1) (11)\\
\enddata
\tablecomments{\tablenotetext{a}{These references correspond to the works that have utilized JWST NIRSpec PRISM/CLEAR spectra of the given source.}}
\tablerefs{(1) \citet{Beiler2024ApJ}, (2) \citet{2024ApJ...976...82T}, (3) \citet{2024AJ....167...19L}, (4) \citet{Beiler2023ApJ}, (5) \citet{2024ApJ...975..162L}, (6) \citet{2024AJ....167....5L}, (7) \citet{2025ApJ...980..230T}, (8) \citet{2024AJ....167..237L}, (9) \citet{2024ApJ...975...31H}, (10) \citet{2025MNRAS.tmpL..66L}, (11) \citet{2025arXiv250905514L}.}
\end{deluxetable*}

\section{Spectral Analysis} \label{sec:spec_analysis}

During the initial visual inspection, we checked the full wavelength range (0.6-5.3\,$\mu$m) of the NIRSpec PRISM/CLEAR spectra in order to identify the overall spectral features characteristic of brown dwarfs. After identifying the candidate sources, we truncated the blue end of each spectrum, removing the portion where the signal-to-noise ratio (S/N) $<$ 5. This decision was motivated by the typically low flux levels of brown dwarfs at short wavelengths, combined with the relatively large flux uncertainties in this spectral region.

Since all brown dwarf samples originate from different programs and were observed for diverse scientific purposes, they exhibit significant variations in their environments, effective temperatures, atmospheric structures, and other physical parameters. Consequently, while our analysis is broadly based on a nested sampling Monte Carlo algorithm, we apply additional source-specific treatments for certain exceptional cases. These special analyses are detailed later in this section.

\subsection{Flux Calibration}

\begin{figure*}[t]
\centering 
\includegraphics[width=0.5\textwidth]{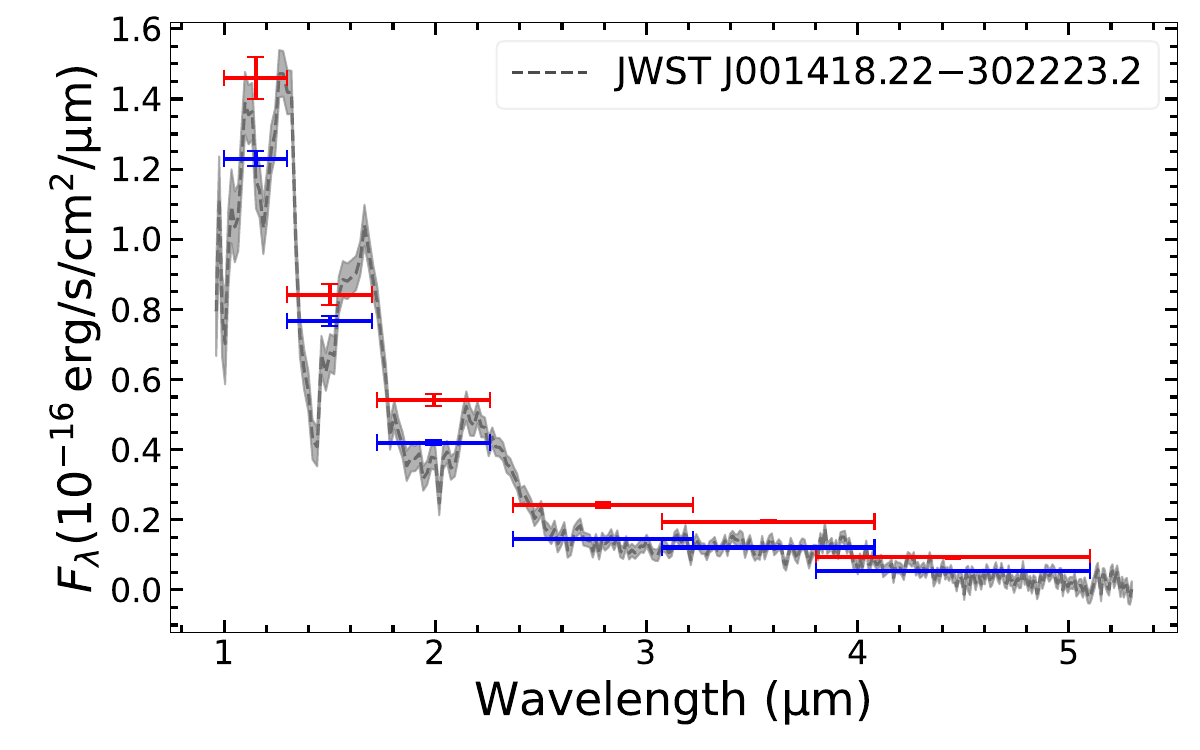}\includegraphics[width=0.5\textwidth]{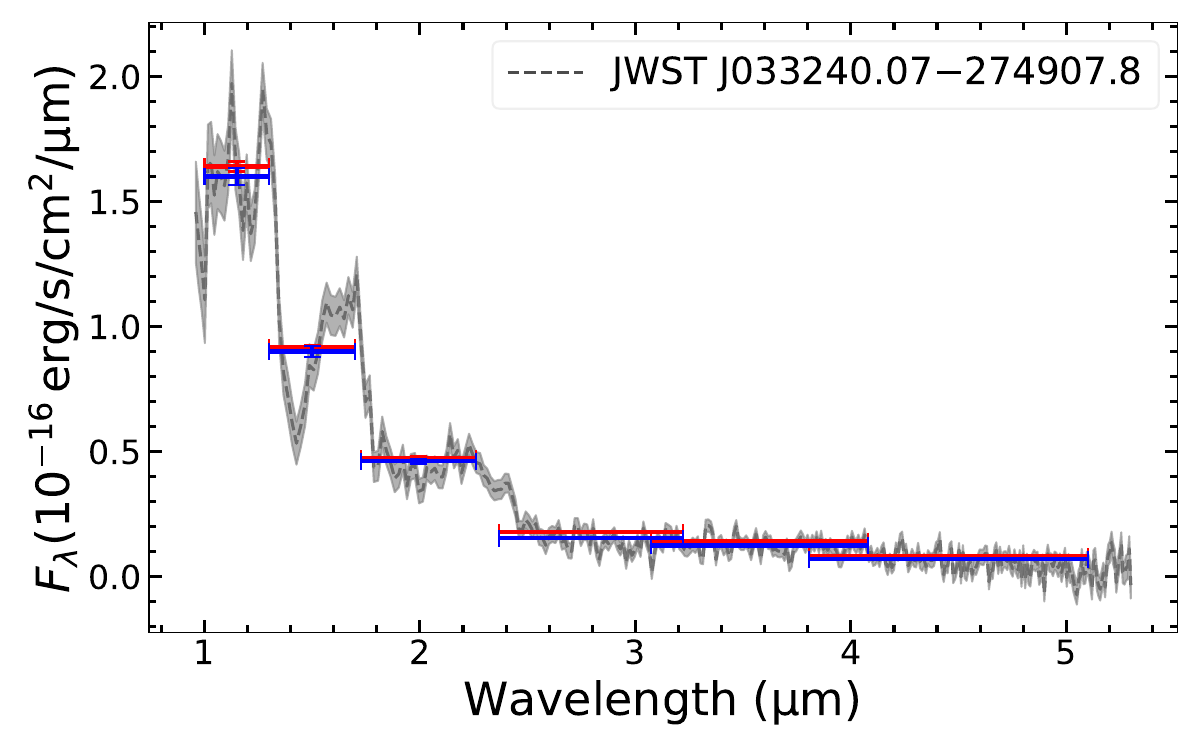}
\caption{Flux calibration check for the two most distant brown dwarf candidates. Red points represent the NIRCam photometric measurements, while blue points show the synthetic photometry obtained by convolving the observed spectra with the corresponding NIRCam filter transmission curves.\\\label{fig:phot_check}} 
\end{figure*}

We note that all of these new brown dwarf candidates, as well as those brown dwarfs identified from the extragalactic deep field surveys (e.g., JADES-GS-BD-9), were observed using the NIRSpec Multi-Object Spectroscopy (MOS) mode. Due to the observational limitations of this mode, combined with the small physical dimensions of the micro shutters (0\farcs20 $\times$ 0\farcs46), not all targets in this mode are perfectly centered within the open shutters \citep{2022A&A...661A..81F}. This misalignment can introduce systematic offsets in the observed spectral flux.

To calibrate the spectral fluxes of such brown dwarf sample, we cross-matched the sources with available NIRCam photometry from several extragalactic deep field surveys \citep{2023ApJS..269...16R,2024A&A...691A.240M,2024ApJS..270....7W}. During initial tests with a cross-matching radius of 0.1--0.2 arcsec, we found cases where the NIRCam photometry was completely inconsistent with our spectra. Therefore, we adopted a radius of 0.1 arcsec for the final cross-match. We then compared the observed NIRCam photometry in the F150W, F200W, F277W, F356W, and F444W filters with synthetic photometry derived from our spectra. A linear fit between the observed and synthetic photometry was used to determine a calibration factor, which was applied to rescale the fluxes.

Figure~\ref{fig:phot_check} illustrates two representative cases encountered during the flux calibration of these sources. In the first case, the overall NIRCam photometry appears systematically higher than the synthetic photometry, which can occur when the target is not centered in the shutter during MOS observations (e.g., JWST J001418.22$-$302223.2). In such cases, we calculate a calibration factor to rescale the fluxes. In the second case, the NIRCam photometry and synthetic photometry are broadly consistent, as shown for JWST J033240.07$-$274907.8.

\subsection{Atmospheric Models}

\begin{deluxetable}{ccccccc}

\tablecaption{Atmosphere Model Parameters \label{tab:model}}

\tablehead{\colhead{Model} & \colhead{$T_\mathrm{eff}$} & \colhead{$\log g$} & \colhead{[M/H]} & \colhead{C/O} & \colhead{$\log K_{zz}$} & \colhead{$\mathrm{[\alpha /Fe]}$} \\ 
\colhead{} & \colhead{(K)} & \colhead{($\mathrm{cm\,s^{-2}}$)} & \colhead{} & \colhead{} & \colhead{($\mathrm{cm^2\,s^{-1}}$)} & \colhead{} } 

\startdata
Sonora Elf Owl & 275--2400 & 3.25--5.5 & $-$1.0 to +1.0 & 0.22--1.12 & 2--8 & ... \\
LOWZ & 500--1600 & 3.5--5.25 & $-$2.5 to +1.0 & 0.1--0.85 & $-$1 to +10 & ... \\
SAND & 700--4000 & 2--6 & $-$2.4 to +0.3 & ... & ... & $-$0.05 to +0.4 \\
\enddata

\tablecomments{The solar carbon-to-oxygen ratio (C/O) is 0.458.}

\end{deluxetable}

Most of the previously identified brown dwarfs are located within 100 pc. In contrast, our newly identified 12 brown dwarf candidates are estimated to be located at distances of several hundred to several thousand (or even over ten thousand, see Section \ref{subsec:distance}) pc, strongly suggesting that they likely belong to the Galactic thick disk or halo population. Given the diversity of our brown dwarf sample, we adopted three distinct atmospheric model grids to determine their physical parameters: Sonora Elf Owl (version 2) \citep{2025RNAAS...9..108W}, LOWZ \citep{Meisner2021ApJ}, and SAND \citep{SAND}. The parameters and parameter ranges explored for these models are summarized in Table \ref{tab:model}.

The Sonora Elf Owl model is a cloud-free atmospheric model optimized for fitting the spectra of cooler T- and Y-type brown dwarfs. Compared to its predecessor \citep[version 1,][]{Mukherjee2024ApJ}, version 2 incorporates improved molecular opacities, including updated abundances of CO$_2$ and PH$_3$ as discussed by \citet{2024ApJ...973...60B}. This model is particularly suited for low-temperature, near-solar metallicity objects \citep{2024ApJ...976...82T,Burgasser2025ApJ}. 

The LOWZ and SAND models are both tailored for studying metal-poor brown dwarfs associated with the Galactic thick disk and halo. LOWZ is a cloud-free atmospheric grid that extends to metallicities as low as [M/H] = $-2.5$, making it particularly effective for fitting spectra of metal-poor subdwarfs \citep{Burgasser2024ApJ}. In contrast, the SAND model incorporates the effects of condensate clouds and emphasizes their role in shaping the spectra of M/L dwarfs. While it also targets metal-poor ultracool dwarfs, SAND provides improved fits for the spectra of metal-poor M-, L-, and T-type subdwarfs where cloud opacity plays a significant role \citep{Burgasser2025ApJ}. Together, these two models complement each other in characterizing the atmospheric properties of brown dwarfs in the thick disk, halo, and globular clusters.

\subsection{Fitting with Atmospheric Models}

To perform atmospheric parameter estimation, we first linearly interpolated each of the three model grids across their native parameter spaces to construct continuous spectral libraries. These interpolated spectra were subsequently convolved with wavelength-dependent Gaussian kernels to match the spectral resolution of the NIRSpec PRISM/CLEAR observations, following the resolution-wavelength relation reported by \citet{2022AA...661A..80J}. 

The posterior probability distributions of the physical parameters were derived using the nested sampling Monte Carlo algorithm \texttt{MLFriends} \citep{untranest2016, ultranest2019}, implemented within the \texttt{UltraNest} framework \citep{ultranest}. The likelihood function adopted in our analysis is defined as:

\begin{equation}
\ln L = -0.5 \sum_i \frac{(F_i - \frac{R^2}{D^2} M_i)^2}{\sigma_i^2+f^2} - \ln\sqrt{2\pi (\sigma_i^2+f^2)},
\end{equation}

where $F_i$, $M_i$, and $\sigma_i$ denote the observed flux, model flux, and flux uncertainty at the $i$-th wavelength point, respectively. The scaling factor $R^2/D^2$, representing the square of the ratio between the object’s radius and distance, was treated as a free parameter in the inference. We adopted flat priors for all parameters, including a logarithmic prior on the scale factor, $\log (R^2/D^2)$, spanning the range from $-10$ to $-30$.

In our analysis, we adopted the standard criteria implemented in \texttt{UltraNest}. Specifically, we used 400 live points, required the fraction of remaining live point weights to fall below $f_\mathrm{live}<0.01$, and adopted the uncertainty in the log-evidence $\Delta \log Z < 0.5$ \citep{2009AIPC.1193..277S,ultranest2019,ultranest}. Once all runs satisfied these conditions, the sampler was terminated and the final evidence values and posterior samples were returned.

To account for possible underestimation of flux uncertainties or additional sources of systematic noise, we introduced a jitter term $f$, which was treated as a free parameter in the fitting. By allowing $f$ to vary, the model can accommodate potential mismatches between the assumed and true noise properties in the observed spectra. The effect of $f$ is particularly pronounced in cases where the reduced chi-square $\chi_r^2$ is significantly greater than 1. In our preliminary tests, we found that introducing $f$ increased the posterior uncertainties compared to the runs without $f$, by about 20–200\% depending on the parameter. This indicates that the inclusion of $f$ yields noticeably broader and thus more realistic posterior uncertainties.

Considering that linear interpolation of model spectra may introduce substantial uncertainties due to the finite grid spacing, we performed leave-one-out validation tests on the LOWZ and SAND models, following the method described in Appendix A of \citet{2024ApJ...976...82T}. The results show that the parameter uncertainties caused by linear interpolation in the LOWZ and SAND models are consistent with those reported for the Sonora Elf Owl grid in \citet{2024ApJ...976...82T}, with most discrepancies being approximately half of the model grid spacing. Therefore, we adopt half of the model grid spacing as one of the final parameter uncertainties \citep{2008ApJ...678.1372C,2024AJ....167..168M,Hurt2024ApJ}. For $\log(R^{2}/D^{2})$, based on our leave-one-out validation test results and the findings of \citet{2024ApJ...976...82T} for the Sonora Elf Owl models, we adopted an uncertainty of 0.05 dex as an additional term.

\subsection{Brown Dwarfs in Star Clusters}

Our sample includes a subset of brown dwarfs located in the IC 348 and Orion Nebula Cluster (ONC) star-forming regions. These spectra are drawn from JWST PID 1228, 1229, and 2770, and the spectral types are primarily in the M to L range \citep{2024AJ....168..230L,2024ApJ...975..162L,2025MNRAS.tmpL..66L}. Analysis indicates that some of the brown dwarf spectra in the ONC exhibit strong nebular emission lines and potential infrared excesses that may originate from surrounding circumstellar disks \citep{2024ApJ...975..162L,2025MNRAS.tmpL..66L}. Moreover, most of these spectra are affected by extinction due to interstellar dust, with several sources exhibiting high levels of extinction.

For this subset of sources, we adopted the SAND atmospheric model exclusively, as it is specifically designed to be better suited for young and potentially dusty environments. To account for the significant dust extinction in these regions, we included and treated the extinction parameter ($A_\mathrm{V}$) as a free parameter during the fitting process. The extinction correction was applied using the wavelength-dependent extinction law ($A_\lambda/A_\mathrm{V}$) from \citet{2024ApJ...964L...3W}, which was empirically derived for the JWST/NIRSpec spectral range (0.6-5.3\,$\mu$m). This law closely aligns with the widely adopted Galactic average extinction curve of \citet{2019ApJ...877..116W,2023ApJ...946...43W}, while providing improved accuracy for cool, low-mass stars and brown dwarfs. Since the \citet{2024ApJ...964L...3W} law is based on a sample of M- and L-type dwarf candidates---comparable to our targets in both spectral type and observational setup---it is particularly appropriate for our sources. 

\section{Results} \label{sec:results}

The best-fit parameters derived from each atmospheric model for all 68 brown dwarfs are summarized in Table~\ref{tab:allsample}, with the first 12 sources corresponding to our newly discovered brown dwarf candidates. If the best-fit $T_\mathrm{eff}$ of a given source reaches the boundary of the model grid, the result is not included in the table. Consequently, for the warmer candidates ($T_\mathrm{eff} > 1600\,\mathrm{K}$), only the fitting results from the Sonora Elf Owl and SAND models are provided. 

We examined the comparison between the best-fit model spectra and the observed spectra for all sources and found that the Sonora Elf Owl and LOWZ models sometimes fail to reproduce the spectral features of brown dwarfs in the L/T transition phase. Therefore, when a source has multiple model fits, we mark the result of a given model with an asterisk (*) in Table \ref{tab:allsample} if its $\chi_r^2$ is about an order of magnitude (i.e., $\sim$10 times) larger than the minimum $\chi_r^2$ obtained among all models for that source. For sources with only a single model fit, we also mark the result with an asterisk if, upon visual inspection, the fit appears unsatisfactory. This mark indicates that the derived parameters from these models should be treated with caution.

\begin{longrotatetable}
\begin{deluxetable}{cccccccccccccccccc}
\tabletypesize{\scriptsize}
\setlength{\tabcolsep}{0.02cm}

\tablecaption{Brown Dwarf Sample: Derived Parameters\label{tab:allsample}}

\tablehead{\colhead{Object Name} & \colhead{SpT$^{a}$} & \colhead{Ref.} & \colhead{Model$^{b}$} & \colhead{$T_\mathrm{eff}$} & \colhead{$\log g$$^{c}$} & \colhead{[M/H]} & \colhead{C/O} & \colhead{$\log K_\mathrm{zz}$} & \colhead{$\alpha$/Fe} & \colhead{$\mathrm{A_v}$} & \colhead{$\log (R^2/D^2)$} & \colhead{$\log f$} & \colhead{$\chi^2_r$} & \colhead{Age} & \colhead{$M$} & \colhead{$R$} & \colhead{$D$} \\ 
\colhead{} & \colhead{} & \colhead{} & \colhead{} & \colhead{(K)} & \colhead{} & \colhead{} & \colhead{} & \colhead{} & \colhead{} & \colhead{(mag)} & \colhead{} & \colhead{} & \colhead{} & \colhead{(Gyr)} & \colhead{($M_\mathrm{jup}$)} & \colhead{($R_\mathrm{jup}$)} & \colhead{(pc)}  } 

\startdata
JWST J001418.22$-$302223.2 & L1 & (0) & Sonora & $2177^{+54}_{-53}$ & $\ddagger3.38^{+0.2}_{-0.16}$ & $-0.54^{+0.28}_{-0.26}$ & $0.64^{+0.12}_{-0.12}$ & $5.53^{+2.81}_{-2.87}$ &  &  & $-24.8^{+0.05}_{-0.05}$ & $-17.76^{+0.05}_{-0.06}$ & 1.6 & $0.001^{+0.001}_{-0.001}$ & $14.2^{+24.4}_{-7.9}$ & $2.77^{+1.90}_{-0.65}$ & $15874^{+6706}_{-6367}$ \\
 &  &  & SAND & $2214^{+53}_{-53}$ & $4.15^{+0.3}_{-0.29}$ & $-0.3^{+0.16}_{-0.14}$ &  &  & $-0.01^{+0.04}_{-0.03}$ &  & $-24.82^{+0.05}_{-0.05}$ & $-17.77^{+0.05}_{-0.06}$ & 1.5 & $0.020^{+0.013}_{-0.017}$ & $15.5^{+7.3}_{-4.6}$ & $1.68^{+0.29}_{-0.24}$ & $9827^{+1607}_{-1643}$ \\
JWST J021651.35$-$050623.6 & T3 & (0) & Sonora & $1269^{+57}_{-57}$ & $\ddagger3.28^{+0.13}_{-0.13}$ & $-0.25^{+0.26}_{-0.26}$ & $0.25^{+0.13}_{-0.12}$ & $6.11^{+1.63}_{-1.63}$ &  &  & $-24.1^{+0.06}_{-0.06}$ & $-17.45^{+0.06}_{-0.07}$ & 1.4 & $0.001^{+0.001}_{-0.001}$ & $2.6^{+3.2}_{-0.4}$ & $1.71^{+0.31}_{-0.06}$ & $4347^{+583}_{-542}$ \\
 &  &  & LOWZ & $1279^{+57}_{-58}$ & $\ddagger3.68^{+0.32}_{-0.28}$ & $-0.45^{+0.15}_{-0.15}$ & $0.36^{+0.24}_{-0.24}$ & $6.54^{+4.17}_{-4.14}$ &  &  & $-24.12^{+0.06}_{-0.06}$ & $-17.45^{+0.06}_{-0.06}$ & 1.4 & $0.004^{+0.008}_{-0.003}$ & $4.4^{+2.8}_{-1.6}$ & $1.52^{+0.16}_{-0.13}$ & $3947^{+479}_{-449}$ \\
 &  &  & SAND & $1262^{+74}_{-57}$ & $4.68^{+0.37}_{-0.28}$ & $-0.6^{+0.14}_{-0.14}$ &  &  & $0.22^{+0.09}_{-0.07}$ &  & $-24.1^{+0.07}_{-0.09}$ & $-17.54^{+0.07}_{-0.1}$ & 1.3 & $0.172^{+0.431}_{-0.083}$ & $21.4^{+14.4}_{-8.4}$ & $1.09^{+0.13}_{-0.14}$ & $2763^{+435}_{-407}$ \\
JWST J021701.94$-$050621.7 & T5$\pm$2 & (0) & Sonora & $739^{+42}_{-39}$ & $5.18^{+0.25}_{-0.38}$ & $-0.27^{+0.29}_{-0.3}$ & $0.42^{+0.18}_{-0.15}$ & $2.73^{+1.34}_{-1.12}$ &  &  & $-23.45^{+0.08}_{-0.09}$ & $-17.47^{+0.08}_{-0.12}$ & 1.3 & $5.166^{+4.834}_{-3.625}$ & $41.7^{+15.9}_{-15.8}$ & $0.84^{+0.11}_{-0.07}$ & $1008^{+157}_{-138}$ \\
 &  &  & LOWZ & $745^{+52}_{-44}$ & $5.08^{+0.18}_{-0.3}$ & $-0.41^{+0.23}_{-0.23}$ & $0.57^{+0.2}_{-0.17}$ & $5.35^{+4.14}_{-4.2}$ &  &  & $-23.45^{+0.11}_{-0.1}$ & $-17.43^{+0.07}_{-0.08}$ & 1.3 & $3.365^{+6.635}_{-2.016}$ & $35.8^{+14.0}_{-10.5}$ & $0.88^{+0.08}_{-0.08}$ & $1056^{+167}_{-147}$ \\
 &  &  & SAND & $739^{+57}_{-53}$ & $5.81^{+0.28}_{-0.28}$ & $0.15^{+0.12}_{-0.12}$ &  &  & $0.0^{+0.06}_{-0.05}$ &  & $-23.47^{+0.08}_{-0.08}$ & $-17.45^{+0.08}_{-0.1}$ & 1.3 & $\dagger10.000^{+0.001}_{-0.001}$ & $\dagger64.9^{+3.1}_{-6.3}$ & $\dagger0.76^{+0.01}_{-0.01}$ & $934^{+88}_{-80}$ \\
JWST J021715.61$-$050857.7 & T5 & (0) & Sonora & $1051^{+50}_{-50}$ & $\ddagger3.27^{+0.13}_{-0.13}$ & $-0.61^{+0.25}_{-0.25}$ & $0.28^{+0.12}_{-0.12}$ & $3.84^{+1.05}_{-1.05}$ &  &  & $-22.92^{+0.05}_{-0.05}$ & $-16.97^{+0.06}_{-0.06}$ & 1.8 & $0.002^{+0.001}_{-0.001}$ & $1.9^{+0.5}_{-0.3}$ & $1.60^{+0.06}_{-0.06}$ & $1044^{+74}_{-70}$ \\
 &  &  & LOWZ & $1022^{+50}_{-50}$ & $5.22^{+0.13}_{-0.14}$ & $-0.05^{+0.13}_{-0.13}$ & $0.59^{+0.15}_{-0.15}$ & $5.55^{+4.02}_{-4.02}$ &  &  & $-22.92^{+0.05}_{-0.05}$ & $-16.85^{+0.05}_{-0.04}$ & 2.2 & $2.324^{+1.869}_{-1.001}$ & $46.6^{+10.0}_{-8.5}$ & $0.85^{+0.05}_{-0.05}$ & $555^{+47}_{-44}$ \\
 &  &  & SAND & $1120^{+50}_{-50}$ & $5.8^{+0.27}_{-0.26}$ & $-0.04^{+0.13}_{-0.13}$ &  &  & $-0.03^{+0.03}_{-0.03}$ &  & $-23.05^{+0.05}_{-0.05}$ & $-16.63^{+0.03}_{-0.03}$ & 3.6 & $\dagger10.000^{+0.001}_{-0.001}$ & $\dagger72.3^{+0.1}_{-2.1}$ & $\dagger0.76^{+0.01}_{-0.01}$ & $577^{+34}_{-33}$ \\
JWST J031347.94$-$671235.0 & T2 & (0) & Sonora & $1380^{+50}_{-50}$ & $5.49^{+0.13}_{-0.13}$ & $0.57^{+0.11}_{-0.11}$ & $0.72^{+0.22}_{-0.22}$ & $2.95^{+1.11}_{-1.11}$ &  &  & $-23.1^{+0.05}_{-0.05}$ & $-16.43^{+0.03}_{-0.02}$ & 13.8 & $6.627^{+3.373}_{-4.690}$ & $74.2^{+1.3}_{-12.4}$ & $0.79^{+0.04}_{-0.01}$ & $635^{+42}_{-44}$ \\
 &  &  & LOWZ & $1192^{+50}_{-50}$ & $5.25^{+0.13}_{-0.13}$ & $0.89^{+0.13}_{-0.13}$ & $0.83^{+0.15}_{-0.15}$ & $9.79^{+4.0}_{-4.02}$ &  &  & $-22.9^{+0.05}_{-0.05}$ & $-16.2^{+0.02}_{-0.02}$ & 50.1 & $1.756^{+1.330}_{-0.712}$ & $50.8^{+10.3}_{-9.0}$ & $0.86^{+0.05}_{-0.04}$ & $548^{+45}_{-41}$ \\
 &  &  & SAND & $1537^{+50}_{-50}$ & $6.0^{+0.25}_{-0.25}$ & $-0.14^{+0.13}_{-0.13}$ &  &  & $0.07^{+0.05}_{-0.05}$ &  & $-23.29^{+0.05}_{-0.05}$ & $-16.11^{+0.02}_{-0.02}$ & 82.5 & $\dagger10.000^{+0.001}_{-0.001}$ & $\dagger76.5^{+0.1}_{-1.0}$ & $\dagger0.80^{+0.01}_{-0.01}$ & $801^{+48}_{-45}$ \\
JWST J033225.12$-$274636.1 & T5 & (0) & Sonora & $1041^{+50}_{-50}$ & $5.06^{+0.14}_{-0.14}$ & $0.1^{+0.25}_{-0.25}$ & $0.49^{+0.12}_{-0.12}$ & $2.44^{+1.04}_{-1.03}$ &  &  & $-22.5^{+0.05}_{-0.05}$ & $-16.27^{+0.02}_{-0.03}$ & 9.3 & $1.162^{+0.937}_{-0.495}$ & $37.1^{+8.7}_{-7.1}$ & $0.91^{+0.05}_{-0.05}$ & $367^{+29}_{-27}$ \\
 &  &  & LOWZ & $1008^{+51}_{-50}$ & $4.54^{+0.28}_{-0.27}$ & $0.21^{+0.14}_{-0.13}$ & $0.56^{+0.15}_{-0.15}$ & $4.25^{+4.02}_{-4.04}$ &  &  & $-22.49^{+0.05}_{-0.05}$ & $-16.15^{+0.02}_{-0.01}$ & 20.2 & $0.214^{+0.280}_{-0.147}$ & $16.4^{+8.6}_{-6.0}$ & $1.10^{+0.10}_{-0.10}$ & $437^{+49}_{-45}$ \\
 &  &  & SAND & $1108^{+50}_{-50}$ & $5.67^{+0.25}_{-0.25}$ & $0.04^{+0.13}_{-0.13}$ &  &  & $-0.02^{+0.03}_{-0.03}$ &  & $-22.59^{+0.05}_{-0.05}$ & $-16.05^{+0.02}_{-0.01}$ & 49.4 & $\dagger10.000^{+0.001}_{-5.518}$ & $\dagger71.2^{+1.0}_{-8.5}$ & $\dagger0.76^{+0.03}_{-0.01}$ & $339^{+22}_{-20}$ \\
JWST J033240.07$-$274907.8 & M7.5 & (0) & Sonora & $2287^{+58}_{-57}$ & $\ddagger3.78^{+0.2}_{-0.18}$ & $-0.92^{+0.27}_{-0.26}$ & $0.27^{+0.13}_{-0.12}$ & $5.62^{+2.82}_{-2.88}$ &  &  & $-24.92^{+0.05}_{-0.05}$ & $-17.64^{+0.05}_{-0.06}$ & 1.5 & $0.002^{+0.002}_{-0.001}$ & $11.7^{+4.6}_{-2.0}$ & $2.14^{+0.46}_{-0.23}$ & $13974^{+2317}_{-2247}$ \\
 &  &  & SAND & $2308^{+59}_{-58}$ & $\ddagger3.78^{+0.44}_{-0.44}$ & $-0.89^{+0.15}_{-0.16}$ &  &  & $0.11^{+0.05}_{-0.04}$ &  & $-24.93^{+0.05}_{-0.05}$ & $-17.65^{+0.05}_{-0.06}$ & 1.5 & $0.002^{+0.022}_{-0.001}$ & $13.5^{+14.1}_{-5.0}$ & $2.13^{+0.87}_{-0.45}$ & $13911^{+4324}_{-3993}$ \\
JWST J065810.66$-$555659.5 & T5.5 & (0) & Sonora & $1051^{+50}_{-50}$ & $4.98^{+0.13}_{-0.13}$ & $0.08^{+0.25}_{-0.25}$ & $0.55^{+0.12}_{-0.12}$ & $2.18^{+1.01}_{-1.01}$ &  &  & $-23.24^{+0.05}_{-0.05}$ & $-17.43^{+0.05}_{-0.04}$ & 2.0 & $0.813^{+0.580}_{-0.317}$ & $32.7^{+7.2}_{-5.9}$ & $0.95^{+0.05}_{-0.05}$ & $898^{+69}_{-69}$ \\
 &  &  & LOWZ & $1042^{+51}_{-50}$ & $4.91^{+0.32}_{-0.28}$ & $0.33^{+0.13}_{-0.13}$ & $0.56^{+0.15}_{-0.15}$ & $2.86^{+4.01}_{-4.01}$ &  &  & $-23.28^{+0.05}_{-0.05}$ & $-17.07^{+0.02}_{-0.03}$ & 4.0 & $0.684^{+1.546}_{-0.457}$ & $30.2^{+16.3}_{-11.8}$ & $0.96^{+0.12}_{-0.11}$ & $945^{+130}_{-120}$ \\
 &  &  & SAND & $1111^{+50}_{-50}$ & $5.63^{+0.25}_{-0.25}$ & $0.1^{+0.13}_{-0.13}$ &  &  & $-0.02^{+0.03}_{-0.03}$ &  & $-23.34^{+0.05}_{-0.05}$ & $-16.77^{+0.02}_{-0.02}$ & 10.9 & $\dagger10.000^{+0.001}_{-6.339}$ & $\dagger71.2^{+1.0}_{-11.6}$ & $\dagger0.76^{+0.05}_{-0.01}$ & $804^{+60}_{-53}$ \\
JWST J091049.86$-$041320.8 & T5 & (0) & Sonora & $963^{+25}_{-25}$ & $5.25^{+0.13}_{-0.13}$ & $0.04^{+0.25}_{-0.25}$ & $0.34^{+0.12}_{-0.12}$ & $2.09^{+1.01}_{-1.0}$ &  &  & $-22.28^{+0.05}_{-0.05}$ & $-16.24^{+0.02}_{-0.02}$ & 9.2 & $3.066^{+2.575}_{-1.195}$ & $48.0^{+9.4}_{-8.1}$ & $0.84^{+0.04}_{-0.04}$ & $262^{+20}_{-19}$ \\
 &  &  & LOWZ & $949^{+25}_{-25}$ & $4.51^{+0.26}_{-0.25}$ & $-0.09^{+0.13}_{-0.13}$ & $0.45^{+0.23}_{-0.23}$ & $4.11^{+4.01}_{-4.01}$ &  &  & $-22.25^{+0.05}_{-0.05}$ & $-16.16^{+0.01}_{-0.02}$ & 15.4 & $0.231^{+0.232}_{-0.159}$ & $15.1^{+7.5}_{-5.0}$ & $1.11^{+0.09}_{-0.09}$ & $336^{+35}_{-33}$ \\
 &  &  & SAND & $1036^{+50}_{-50}$ & $5.4^{+0.25}_{-0.25}$ & $-0.08^{+0.13}_{-0.13}$ &  &  & $-0.0^{+0.05}_{-0.05}$ &  & $-22.37^{+0.05}_{-0.05}$ & $-16.03^{+0.02}_{-0.02}$ & 28.7 & $5.519^{+4.481}_{-3.808}$ & $61.1^{+9.1}_{-19.1}$ & $0.79^{+0.09}_{-0.03}$ & $273^{+26}_{-25}$ \\
JWST J141916.32$+$525258.8 & L3 & (0) & Sonora* & $1608^{+54}_{-51}$ & $\ddagger3.26^{+0.13}_{-0.13}$ & $0.7^{+0.17}_{-0.17}$ & $0.76^{+0.22}_{-0.22}$ & $6.32^{+1.6}_{-2.06}$ &  &  & $-22.81^{+0.05}_{-0.05}$ & $-15.5^{+0.02}_{-0.02}$ & 728.7 & $0.001^{+0.001}_{-0.001}$ & $9.0^{+11.5}_{-5.9}$ & $2.29^{+0.93}_{-0.49}$ & $1317^{+394}_{-391}$ \\
 &  &  & SAND & $1664^{+50}_{-50}$ & $4.79^{+0.25}_{-0.25}$ & $-0.88^{+0.13}_{-0.13}$ &  &  & $0.08^{+0.05}_{-0.05}$ &  & $-22.83^{+0.05}_{-0.05}$ & $-16.01^{+0.02}_{-0.01}$ & 57.6 & $0.142^{+0.218}_{-0.073}$ & $28.4^{+12.6}_{-8.5}$ & $1.10^{+0.11}_{-0.11}$ & $646^{+78}_{-70}$ \\
JWST J141919.36$+$525006.9 & M7.5 & (0) & Sonora & $2075^{+70}_{-66}$ & $\ddagger3.3^{+0.15}_{-0.13}$ & $0.97^{+0.15}_{-0.17}$ & $0.55^{+0.13}_{-0.16}$ & $3.32^{+1.44}_{-1.35}$ &  &  & $-24.01^{+0.06}_{-0.07}$ & $-16.53^{+0.03}_{-0.05}$ & 1.9 & $0.001^{+0.001}_{-0.001}$ & $15.0^{+21.4}_{-9.8}$ & $2.80^{+1.70}_{-0.78}$ & $6427^{+2911}_{-2745}$ \\
 &  &  & SAND & $1632^{+178}_{-64}$ & $\ddagger3.66^{+0.25}_{-0.26}$ & $-1.3^{+0.16}_{-0.13}$ &  &  & $0.17^{+0.05}_{-0.04}$ &  & $-23.71^{+0.06}_{-0.13}$ & $-17.14^{+0.27}_{-1.79}$ & 1.2 & $0.002^{+0.004}_{-0.001}$ & $5.8^{+3.5}_{-1.7}$ & $1.68^{+0.19}_{-0.17}$ & $2710^{+436}_{-384}$ \\
JWST J220249.96$+$185059.6 & T2 & (0) & Sonora & $1550^{+54}_{-53}$ & $5.34^{+0.17}_{-0.23}$ & $0.65^{+0.12}_{-0.12}$ & $0.69^{+0.22}_{-0.22}$ & $4.78^{+1.52}_{-1.53}$ &  &  & $-23.95^{+0.05}_{-0.06}$ & $-18.49^{+0.72}_{-1.02}$ & 1.1 & $1.338^{+6.559}_{-0.731}$ & $61.8^{+14.6}_{-14.9}$ & $0.86^{+0.08}_{-0.06}$ & $1839^{+193}_{-185}$ \\
 &  &  & LOWZ & $1457^{+55}_{-62}$ & $5.22^{+0.13}_{-0.13}$ & $0.73^{+0.15}_{-0.15}$ & $0.63^{+0.25}_{-0.15}$ & $4.4^{+6.12}_{-4.06}$ &  &  & $-23.93^{+0.07}_{-0.06}$ & $-17.45^{+0.06}_{-0.07}$ & 1.4 & $0.923^{+0.711}_{-0.368}$ & $50.9^{+10.9}_{-8.8}$ & $0.89^{+0.05}_{-0.05}$ & $1854^{+185}_{-167}$ \\
 &  &  & SAND & $1491^{+51}_{-58}$ & $5.76^{+0.26}_{-0.27}$ & $0.28^{+0.1}_{-0.12}$ &  &  & $0.0^{+0.05}_{-0.05}$ &  & $-23.98^{+0.06}_{-0.05}$ & $-17.4^{+0.05}_{-0.06}$ & 1.6 & $\dagger10.000^{+0.001}_{-3.229}$ & $\dagger75.4^{+1.0}_{-0.1}$ & $\dagger0.80^{+0.01}_{-0.01}$ & $1773^{+116}_{-109}$ \\
 \hline \\
2MASS J00470038$+$6803543 & L7 & (4) & Sonora* & $1027^{+93}_{-78}$ & $\ddagger3.26^{+0.13}_{-0.13}$ & $0.97^{+0.15}_{-0.16}$ & $0.23^{+0.12}_{-0.12}$ & $8.6^{+0.59}_{-0.81}$ &  &  & $-19.1^{+0.12}_{-0.15}$ & $-12.09^{+0.02}_{-0.02}$ & 15747.9 & $0.002^{+0.001}_{-0.001}$ & $1.9^{+0.5}_{-0.3}$ & $1.59^{+0.07}_{-0.07}$ & $13^{+2}_{-2}$ \\
 &  &  & LOWZ* & $1521^{+59}_{-52}$ & $4.48^{+0.25}_{-0.27}$ & $0.94^{+0.13}_{-0.15}$ & $0.85^{+0.15}_{-0.15}$ & $9.9^{+4.0}_{-4.0}$ &  &  & $-19.77^{+0.06}_{-0.06}$ & $-12.16^{+0.02}_{-0.01}$ & 12669.7 & $0.077^{+0.064}_{-0.055}$ & $17.2^{+8.0}_{-5.9}$ & $1.21^{+0.12}_{-0.11}$ & $21^{+3}_{-2}$ \\
 &  &  & SAND & $1206^{+50}_{-50}$ & $4.0^{+0.25}_{-0.25}$ & $-0.71^{+0.13}_{-0.13}$ &  &  & $0.21^{+0.05}_{-0.05}$ &  & $-19.2^{+0.05}_{-0.05}$ & $-12.96^{+0.01}_{-0.02}$ & 284.1 & $0.015^{+0.025}_{-0.009}$ & $7.0^{+3.8}_{-2.4}$ & $1.35^{+0.11}_{-0.11}$ & $12^{+1}_{-1}$ \\
WISE J024714.52$+$372523.5 & T8 & (5) & Sonora & $647^{+25}_{-25}$ & $4.48^{+0.13}_{-0.13}$ & $-0.13^{+0.25}_{-0.25}$ & $0.46^{+0.12}_{-0.11}$ & $2.02^{+1.0}_{-1.0}$ &  &  & $-19.78^{+0.05}_{-0.05}$ & $-14.31^{+0.02}_{-0.02}$ & 10.7 & $0.642^{+0.244}_{-0.323}$ & $13.3^{+3.3}_{-2.6}$ & $1.07^{+0.04}_{-0.04}$ & $19^{+1}_{-1}$ \\
 &  &  & LOWZ & $630^{+26}_{-25}$ & $5.23^{+0.13}_{-0.13}$ & $-0.05^{+0.13}_{-0.13}$ & $0.82^{+0.15}_{-0.15}$ & $6.3^{+4.01}_{-4.01}$ &  &  & $-19.74^{+0.05}_{-0.05}$ & $-14.07^{+0.01}_{-0.02}$ & 31.6 & $\dagger10.000^{+0.001}_{-4.043}$ & $\dagger43.0^{+5.2}_{-7.0}$ & $\dagger0.82^{+0.04}_{-0.02}$ & $14^{+1}_{-1}$ \\
WISEPA J031325.96$+$780744.2 & T8.5 & (6) & Sonora & $558^{+12}_{-12}$ & $\ddagger3.27^{+0.13}_{-0.13}$ & $-0.4^{+0.25}_{-0.25}$ & $0.31^{+0.12}_{-0.12}$ & $3.9^{+1.0}_{-1.01}$ &  &  & $-19.04^{+0.05}_{-0.05}$ & $-13.98^{+0.01}_{-0.02}$ & 24.7 & $0.010^{+0.005}_{-0.003}$ & $1.3^{+0.4}_{-0.3}$ & $1.33^{+0.03}_{-0.03}$ & $10^{+1}_{-1}$ \\
 &  &  & LOWZ & $528^{+26}_{-25}$ & $5.23^{+0.13}_{-0.13}$ & $-0.08^{+0.13}_{-0.13}$ & $0.81^{+0.15}_{-0.15}$ & $6.38^{+4.01}_{-4.01}$ &  &  & $-19.01^{+0.05}_{-0.06}$ & $-13.64^{+0.02}_{-0.02}$ & 108.1 & $\dagger10.000^{+0.001}_{-0.001}$ & $\dagger39.8^{+6.3}_{-5.2}$ & $\dagger0.83^{+0.02}_{-0.02}$ & $6^{+1}_{-1}$ \\
JADES-GS-BD-9 & T5-T6 & (3) & Sonora & $866^{+48}_{-43}$ & $4.58^{+0.4}_{-0.55}$ & $-0.48^{+0.31}_{-0.32}$ & $0.36^{+0.17}_{-0.15}$ & $4.65^{+1.86}_{-1.89}$ &  &  & $-23.93^{+0.09}_{-0.08}$ & $-17.79^{+0.08}_{-0.14}$ & 1.2 & $0.342^{+1.467}_{-0.289}$ & $16.8^{+17.0}_{-9.2}$ & $1.07^{+0.16}_{-0.16}$ & $2233^{+412}_{-375}$ \\
 &  &  & LOWZ & $884^{+43}_{-44}$ & $5.0^{+0.31}_{-0.41}$ & $-0.46^{+0.17}_{-0.16}$ & $0.62^{+0.19}_{-0.17}$ & $6.58^{+4.12}_{-4.15}$ &  &  & $-23.96^{+0.08}_{-0.08}$ & $-17.76^{+0.08}_{-0.1}$ & 1.3 & $1.456^{+5.320}_{-1.069}$ & $32.4^{+22.2}_{-13.7}$ & $0.92^{+0.12}_{-0.13}$ & $1978^{+337}_{-315}$ \\
 &  &  & SAND & $895^{+78}_{-74}$ & $5.89^{+0.26}_{-0.28}$ & $0.08^{+0.14}_{-0.16}$ &  &  & $-0.01^{+0.04}_{-0.03}$ &  & $-24.04^{+0.09}_{-0.1}$ & $-17.72^{+0.07}_{-0.09}$ & 1.3 & $\dagger10.000^{+0.001}_{-0.001}$ & $\dagger69.1^{+1.0}_{-3.1}$ & $\dagger0.76^{+0.01}_{-0.01}$ & $1801^{+213}_{-185}$ \\
IC 348 MM 51 & ? &  & SAND & $2823^{+60}_{-59}$ & $\ddagger2.89^{+0.28}_{-0.4}$ & $-1.74^{+0.22}_{-0.31}$ &  &  & $0.32^{+0.03}_{-0.03}$ & $4.33^{+0.07}_{-0.08}$ & $-22.76^{+0.05}_{-0.05}$ & $-15.15^{+0.02}_{-0.02}$ & 734.1 & $\dagger0.001^{+0.001}_{-0.001}$ & $\dagger78.6^{+0.1}_{-10.5}$ & $\dagger7.92^{+0.01}_{-0.78}$ & $4300^{+363}_{-334}$ \\
2MASS J034807.72$-$602227.0 & T7 & (7) & Sonora & $859^{+25}_{-25}$ & $3.75^{+0.13}_{-0.13}$ & $-0.18^{+0.25}_{-0.25}$ & $0.45^{+0.12}_{-0.12}$ & $2.01^{+1.0}_{-1.0}$ &  &  & $-19.16^{+0.05}_{-0.05}$ & $-13.16^{+0.02}_{-0.02}$ & 10334.1 & $0.015^{+0.009}_{-0.006}$ & $3.9^{+1.1}_{-0.9}$ & $1.34^{+0.04}_{-0.04}$ & $12^{+1}_{-1}$ \\
 &  &  & LOWZ & $882^{+26}_{-25}$ & $4.05^{+0.26}_{-0.25}$ & $0.12^{+0.13}_{-0.13}$ & $0.46^{+0.23}_{-0.23}$ & $2.45^{+4.02}_{-4.03}$ &  &  & $-19.25^{+0.05}_{-0.06}$ & $-13.0^{+0.02}_{-0.01}$ & 20495.8 & $0.040^{+0.063}_{-0.024}$ & $6.8^{+3.6}_{-2.6}$ & $1.25^{+0.08}_{-0.08}$ & $12^{+1}_{-1}$ \\
 &  &  & SAND & $937^{+51}_{-50}$ & $5.82^{+0.25}_{-0.25}$ & $0.11^{+0.1}_{-0.11}$ &  &  & $0.01^{+0.05}_{-0.05}$ &  & $-19.32^{+0.05}_{-0.05}$ & $-12.89^{+0.01}_{-0.02}$ & 117339.1 & $\dagger10.000^{+0.001}_{-0.001}$ & $\dagger69.1^{+1.0}_{-3.1}$ & $\dagger0.76^{+0.01}_{-0.01}$ & $8^{+1}_{-1}$ \\
2MASS J03552337$+$1133437 & L3 & (8) & Sonora* & $1588^{+54}_{-62}$ & $\ddagger3.27^{+0.13}_{-0.13}$ & $0.73^{+0.2}_{-0.17}$ & $0.69^{+0.22}_{-0.22}$ & $6.37^{+1.62}_{-2.16}$ &  &  & $-19.28^{+0.06}_{-0.06}$ & $-11.51^{+0.01}_{-0.02}$ & 18786.4 & $0.001^{+0.001}_{-0.001}$ & $7.6^{+11.3}_{-4.4}$ & $2.17^{+0.91}_{-0.37}$ & $22^{+6}_{-6}$ \\
 &  &  & SAND & $1623^{+50}_{-50}$ & $4.0^{+0.25}_{-0.25}$ & $-0.54^{+0.13}_{-0.13}$ &  &  & $-0.01^{+0.03}_{-0.03}$ &  & $-19.14^{+0.05}_{-0.05}$ & $-12.45^{+0.01}_{-0.02}$ & 339.9 & $0.008^{+0.018}_{-0.005}$ & $8.5^{+3.9}_{-2.6}$ & $1.48^{+0.15}_{-0.14}$ & $12^{+1}_{-1}$ \\
WISE J035934.06$-$540154.6 & Y0 & (9) & Sonora & $430^{+12}_{-12}$ & $\ddagger3.26^{+0.13}_{-0.13}$ & $-0.73^{+0.25}_{-0.25}$ & $0.34^{+0.12}_{-0.12}$ & $7.13^{+0.52}_{-0.52}$ &  &  & $-19.57^{+0.05}_{-0.05}$ & $-14.93^{+0.02}_{-0.02}$ & 4.9 & $0.022^{+0.018}_{-0.008}$ & $1.2^{+0.4}_{-0.1}$ & $1.26^{+0.03}_{-0.03}$ & $17^{+1}_{-1}$ \\
WISE J043052.92$+$463331.6 & T8 & (5) & Sonora & $551^{+12}_{-12}$ & $\ddagger3.42^{+0.19}_{-0.14}$ & $-0.85^{+0.25}_{-0.25}$ & $0.22^{+0.12}_{-0.12}$ & $6.0^{+1.51}_{-1.51}$ &  &  & $-19.51^{+0.05}_{-0.05}$ & $-14.55^{+0.02}_{-0.02}$ & 3.9 & $0.016^{+0.014}_{-0.007}$ & $1.8^{+0.7}_{-0.5}$ & $1.30^{+0.03}_{-0.03}$ & $17^{+1}_{-1}$ \\
 &  &  & LOWZ & $511^{+25}_{-25}$ & $5.2^{+0.13}_{-0.15}$ & $-0.43^{+0.13}_{-0.13}$ & $0.53^{+0.23}_{-0.23}$ & $7.3^{+4.01}_{-4.01}$ &  &  & $-19.4^{+0.05}_{-0.05}$ & $-14.09^{+0.02}_{-0.02}$ & 20.7 & $\dagger10.000^{+0.001}_{-0.001}$ & $\dagger38.8^{+5.2}_{-6.2}$ & $\dagger0.84^{+0.03}_{-0.02}$ & $10^{+1}_{-1}$ \\
ONC J083.77378$-$05.37957 & $>$M9 & (10) & SAND & $1835^{+141}_{-62}$ & $3.36^{+0.26}_{-0.26}$ & $-1.88^{+0.18}_{-0.18}$ &  &  & $0.28^{+0.06}_{-0.06}$ & $6.45^{+0.22}_{-0.32}$ & $-21.69^{+0.05}_{-0.07}$ & $-14.85^{+0.02}_{-0.01}$ & 103.5 & $0.001^{+0.001}_{-0.001}$ & $6.0^{+17.0}_{-1.8}$ & $1.91^{+1.51}_{-0.11}$ & $302^{+105}_{-114}$ \\
ONC J083.78378$-$05.37705 & M8 & (11) & SAND & $2726^{+51}_{-52}$ & $3.69^{+0.28}_{-0.27}$ & $-0.71^{+0.15}_{-0.16}$ &  &  & $0.4^{+0.03}_{-0.03}$ & $6.06^{+0.05}_{-0.04}$ & $-21.36^{+0.05}_{-0.05}$ & $-14.21^{+0.02}_{-0.02}$ & 278.4 & $0.001^{+0.007}_{-0.001}$ & $27.1^{+28.8}_{-5.9}$ & $3.42^{+2.71}_{-0.71}$ & $369^{+162}_{-158}$ \\
ONC J083.78472$-$05.38936 & M8 & (11) & SAND & $1830^{+53}_{-52}$ & $3.37^{+0.25}_{-0.25}$ & $-0.79^{+0.13}_{-0.13}$ &  &  & $0.04^{+0.05}_{-0.05}$ & $3.76^{+0.12}_{-0.13}$ & $-22.02^{+0.05}_{-0.05}$ & $-15.23^{+0.02}_{-0.02}$ & 22.8 & $0.001^{+0.001}_{-0.001}$ & $6.2^{+17.1}_{-2.0}$ & $1.91^{+1.54}_{-0.11}$ & $442^{+155}_{-169}$ \\
ONC J083.78535$-$05.40039 & M3.5-L0 & (12) & SAND & $2611^{+51}_{-51}$ & $3.11^{+0.26}_{-0.26}$ & $0.27^{+0.1}_{-0.11}$ &  &  & $0.01^{+0.05}_{-0.05}$ & $6.96^{+0.08}_{-0.06}$ & $-21.58^{+0.05}_{-0.05}$ & $-14.55^{+0.02}_{-0.01}$ & 379.6 & $\dagger0.001^{+0.001}_{-0.001}$ & $\dagger15.7^{+55.5}_{-2.1}$ & $\dagger3.03^{+4.35}_{-0.19}$ & $423^{+277}_{-273}$ \\
ONC J083.78972$-$05.33420 & M9-L2 & (1) & SAND & $1895^{+52}_{-51}$ & $3.25^{+0.25}_{-0.25}$ & $-0.73^{+0.13}_{-0.13}$ &  &  & $0.03^{+0.05}_{-0.05}$ & $5.29^{+0.08}_{-0.07}$ & $-21.71^{+0.05}_{-0.05}$ & $-15.16^{+0.02}_{-0.02}$ & 7.0 & $\dagger0.001^{+0.001}_{-0.001}$ & $\dagger5.2^{+20.5}_{-1.0}$ & $\dagger1.91^{+1.73}_{-0.01}$ & $310^{+133}_{-113}$ \\
ONC J083.80006$-$05.38467 & ? &  & SAND & $2522^{+61}_{-60}$ & $3.36^{+0.33}_{-0.32}$ & $0.18^{+0.13}_{-0.15}$ &  &  & $0.08^{+0.07}_{-0.06}$ & $4.73^{+0.2}_{-0.2}$ & $-21.9^{+0.05}_{-0.05}$ & $-14.56^{+0.02}_{-0.02}$ & 75.0 & $\dagger0.001^{+0.001}_{-0.001}$ & $\dagger14.6^{+31.3}_{-5.2}$ & $\dagger2.84^{+2.45}_{-0.39}$ & $570^{+235}_{-228}$ \\
ONC J083.80445$-$05.34607 & M7.5 & (1) & SAND* & $1856^{+57}_{-60}$ & $3.44^{+0.26}_{-0.26}$ & $-1.53^{+0.13}_{-0.14}$ &  &  & $0.21^{+0.05}_{-0.05}$ & $1.93^{+0.21}_{-0.21}$ & $-21.36^{+0.06}_{-0.05}$ & $-13.95^{+0.01}_{-0.02}$ & 1376.3 & $0.001^{+0.001}_{-0.001}$ & $6.6^{+15.8}_{-2.4}$ & $1.91^{+1.46}_{-0.11}$ & $206^{+75}_{-79}$ \\
ONC J083.81412$-$05.34651 & M9-L2 & (1) & SAND & $2532^{+51}_{-51}$ & $3.5^{+0.26}_{-0.26}$ & $-0.61^{+0.17}_{-0.15}$ &  &  & $0.38^{+0.03}_{-0.04}$ & $6.28^{+0.06}_{-0.06}$ & $-21.68^{+0.05}_{-0.05}$ & $-14.69^{+0.01}_{-0.02}$ & 29.0 & $0.001^{+0.001}_{-0.001}$ & $16.4^{+30.7}_{-4.8}$ & $2.84^{+2.54}_{-0.32}$ & $446^{+190}_{-186}$ \\
COUP 714 & M8 & (1) & SAND & $2615^{+52}_{-52}$ & $3.61^{+0.45}_{-0.28}$ & $-0.18^{+0.13}_{-0.15}$ &  &  & $0.11^{+0.04}_{-0.04}$ & $2.34^{+0.11}_{-0.09}$ & $-20.84^{+0.05}_{-0.05}$ & $-13.64^{+0.02}_{-0.01}$ & 325.4 & $0.001^{+0.008}_{-0.001}$ & $20.5^{+28.3}_{-4.9}$ & $2.93^{+2.56}_{-0.51}$ & $175^{+77}_{-79}$ \\
WISE J053516.80$-$750024.9 & $\geq$Y1: & (9) & Sonora & $397^{+12}_{-13}$ & $\ddagger3.25^{+0.13}_{-0.13}$ & $-0.92^{+0.25}_{-0.25}$ & $0.42^{+0.12}_{-0.12}$ & $5.65^{+1.52}_{-1.51}$ &  &  & $-19.38^{+0.05}_{-0.05}$ & $-14.84^{+0.02}_{-0.02}$ & 5.0 & $\dagger0.027^{+0.030}_{-0.014}$ & $\dagger1.1^{+0.4}_{-0.1}$ & $\dagger1.25^{+0.05}_{-0.03}$ & $14^{+1}_{-1}$ \\
ONC J083.81892$-$05.44293 & M8.75 & (13) & SAND & $2598^{+50}_{-50}$ & $3.44^{+0.27}_{-0.27}$ & $0.04^{+0.14}_{-0.14}$ &  &  & $0.01^{+0.05}_{-0.05}$ & $2.96^{+0.05}_{-0.05}$ & $-21.06^{+0.05}_{-0.05}$ & $-13.94^{+0.01}_{-0.02}$ & 712.0 & $0.001^{+0.001}_{-0.001}$ & $18.2^{+33.6}_{-4.6}$ & $3.03^{+2.72}_{-0.27}$ & $233^{+102}_{-101}$ \\
ONC J083.82316$-$05.43600 & ? &  & SAND & $2474^{+50}_{-50}$ & $3.36^{+0.3}_{-0.27}$ & $-1.34^{+0.12}_{-0.11}$ &  &  & $0.4^{+0.03}_{-0.03}$ & $5.77^{+0.05}_{-0.05}$ & $-21.13^{+0.05}_{-0.05}$ & $-14.14^{+0.01}_{-0.02}$ & 974.8 & $\dagger0.001^{+0.001}_{-0.001}$ & $\dagger12.8^{+30.4}_{-3.4}$ & $\dagger2.64^{+2.43}_{-0.30}$ & $220^{+91}_{-96}$ \\
ONC J083.82323$-$05.37439 & M8 & (1) & SAND & $2676^{+58}_{-68}$ & $3.55^{+0.37}_{-0.4}$ & $-0.77^{+0.38}_{-0.4}$ &  &  & $0.39^{+0.03}_{-0.04}$ & $5.66^{+0.13}_{-0.14}$ & $-21.28^{+0.06}_{-0.05}$ & $-13.66^{+0.02}_{-0.02}$ & 181.9 & $0.001^{+0.007}_{-0.001}$ & $25.2^{+36.2}_{-7.2}$ & $3.37^{+3.19}_{-0.75}$ & $328^{+172}_{-164}$ \\
ONC J083.82387$-$05.36243 & M9-L2 & (1) & SAND & $1852^{+59}_{-82}$ & $3.43^{+0.27}_{-0.26}$ & $-1.36^{+0.2}_{-0.13}$ &  &  & $0.18^{+0.03}_{-0.04}$ & $4.01^{+0.26}_{-0.31}$ & $-21.5^{+0.07}_{-0.05}$ & $-14.41^{+0.02}_{-0.02}$ & 101.4 & $0.001^{+0.001}_{-0.001}$ & $6.6^{+16.6}_{-2.4}$ & $1.91^{+1.53}_{-0.12}$ & $243^{+86}_{-91}$ \\
ONC J083.82400$-$05.32450 & M8 & (1) & SAND & $2656^{+50}_{-50}$ & $3.83^{+0.26}_{-0.26}$ & $-0.76^{+0.13}_{-0.13}$ &  &  & $0.4^{+0.03}_{-0.03}$ & $6.86^{+0.02}_{-0.03}$ & $-20.77^{+0.05}_{-0.05}$ & $-13.99^{+0.01}_{-0.02}$ & 170.0 & $0.002^{+0.010}_{-0.001}$ & $23.1^{+8.8}_{-4.0}$ & $2.80^{+1.11}_{-0.50}$ & $154^{+43}_{-42}$ \\
ONC J083.82520$-$05.36140 & M8.75 & (1) & SAND & $2596^{+53}_{-51}$ & $3.46^{+0.27}_{-0.26}$ & $0.11^{+0.12}_{-0.11}$ &  &  & $0.04^{+0.05}_{-0.05}$ & $3.8^{+0.09}_{-0.08}$ & $-21.01^{+0.05}_{-0.05}$ & $-13.9^{+0.01}_{-0.02}$ & 297.9 & $0.001^{+0.001}_{-0.001}$ & $18.0^{+32.1}_{-4.4}$ & $3.03^{+2.59}_{-0.35}$ & $217^{+97}_{-90}$ \\
ONC J083.82622$-$05.36372 & M8.5 & (1) & SAND & $1987^{+54}_{-56}$ & $3.25^{+0.25}_{-0.25}$ & $-2.04^{+0.13}_{-0.13}$ &  &  & $0.32^{+0.03}_{-0.03}$ & $4.5^{+0.1}_{-0.1}$ & $-20.98^{+0.05}_{-0.05}$ & $-13.97^{+0.01}_{-0.02}$ & 274.5 & $\dagger0.001^{+0.001}_{-0.001}$ & $\dagger5.2^{+23.8}_{-1.0}$ & $\dagger2.02^{+1.88}_{-0.11}$ & $142^{+62}_{-62}$ \\
ONC J083.83180$-$05.40876 & M7.5 & (11) & SAND & $1870^{+75}_{-72}$ & $3.31^{+0.26}_{-0.26}$ & $-1.64^{+0.15}_{-0.16}$ &  &  & $0.23^{+0.05}_{-0.05}$ & $4.12^{+0.27}_{-0.29}$ & $-20.93^{+0.06}_{-0.07}$ & $-13.65^{+0.02}_{-0.02}$ & 702.7 & $0.001^{+0.001}_{-0.001}$ & $5.8^{+19.6}_{-1.6}$ & $1.92^{+1.70}_{-0.10}$ & $127^{+50}_{-58}$ \\
ONC J083.83189$-$05.41171 & ? &  & SAND & $3004^{+66}_{-66}$ & $3.67^{+0.38}_{-0.36}$ & $-1.33^{+0.28}_{-0.33}$ &  &  & $0.3^{+0.08}_{-0.11}$ & $3.91^{+0.1}_{-0.15}$ & $-21.44^{+0.05}_{-0.06}$ & $-13.79^{+0.01}_{-0.02}$ & 518.7 & $0.003^{+0.004}_{-0.002}$ & $64.0^{+14.6}_{-15.0}$ & $5.57^{+2.35}_{-2.12}$ & $656^{+272}_{-266}$ \\
ONC J083.83198$-$05.37616 & M7.25 & (1) & SAND & $2596^{+53}_{-52}$ & $2.9^{+0.28}_{-0.43}$ & $-1.96^{+0.18}_{-0.3}$ &  &  & $0.38^{+0.03}_{-0.03}$ & $5.71^{+0.08}_{-0.07}$ & $-20.78^{+0.05}_{-0.05}$ & $-13.48^{+0.02}_{-0.01}$ & 1210.2 & $\dagger0.001^{+0.001}_{-0.001}$ & $\dagger21.0^{+57.6}_{-9.4}$ & $\dagger3.39^{+4.52}_{-0.76}$ & $190^{+124}_{-115}$ \\
ONC J083.83626$-$05.41327 & ? &  & SAND & $2169^{+345}_{-285}$ & $3.08^{+1.75}_{-0.37}$ & $-0.82^{+0.26}_{-0.23}$ &  &  & $0.1^{+0.3}_{-0.1}$ & $7.83^{+1.38}_{-0.54}$ & $-21.56^{+0.14}_{-0.07}$ & $-14.3^{+0.02}_{-0.02}$ & 493.9 & $\dagger0.001^{+0.003}_{-0.001}$ & $\dagger6.5^{+32.1}_{-4.4}$ & $\dagger1.91^{+1.21}_{-0.28}$ & $263^{+100}_{-91}$ \\
2MASS J06420559$+$4101599 & L9 & (14) & Sonora* & $1490^{+54}_{-59}$ & $\ddagger3.27^{+0.13}_{-0.13}$ & $0.56^{+0.15}_{-0.12}$ & $0.69^{+0.22}_{-0.22}$ & $6.98^{+1.52}_{-1.53}$ &  &  & $-20.17^{+0.06}_{-0.06}$ & $-12.71^{+0.02}_{-0.01}$ & 4718.8 & $0.001^{+0.001}_{-0.001}$ & $4.2^{+11.3}_{-1.3}$ & $1.88^{+0.93}_{-0.17}$ & $52^{+14}_{-12}$ \\
 &  &  & LOWZ* & $1400^{+50}_{-50}$ & $\ddagger3.51^{+0.25}_{-0.25}$ & $0.24^{+0.13}_{-0.13}$ & $0.84^{+0.15}_{-0.15}$ & $9.92^{+4.0}_{-4.0}$ &  &  & $-20.04^{+0.05}_{-0.05}$ & $-12.87^{+0.02}_{-0.02}$ & 2600.0 & $0.002^{+0.003}_{-0.001}$ & $3.9^{+2.2}_{-1.0}$ & $1.66^{+0.12}_{-0.13}$ & $39^{+4}_{-4}$ \\
 &  &  & SAND & $1231^{+50}_{-50}$ & $4.3^{+0.25}_{-0.25}$ & $-0.66^{+0.13}_{-0.13}$ &  &  & $0.18^{+0.03}_{-0.03}$ &  & $-19.72^{+0.05}_{-0.05}$ & $-13.5^{+0.01}_{-0.02}$ & 121.4 & $0.054^{+0.077}_{-0.036}$ & $11.8^{+5.7}_{-3.9}$ & $1.23^{+0.10}_{-0.10}$ & $20^{+2}_{-2}$ \\
2MASS J06462756$+$7935045 & L9 & (15) & Sonora* & $1500^{+52}_{-52}$ & $\ddagger3.29^{+0.14}_{-0.13}$ & $0.5^{+0.26}_{-0.25}$ & $0.74^{+0.22}_{-0.22}$ & $6.91^{+1.51}_{-1.52}$ &  &  & $-20.21^{+0.05}_{-0.06}$ & $-12.92^{+0.01}_{-0.02}$ & 1412.1 & $0.001^{+0.001}_{-0.001}$ & $3.8^{+11.1}_{-0.8}$ & $1.83^{+0.93}_{-0.11}$ & $53^{+12}_{-14}$ \\
 &  &  & LOWZ* & $1399^{+50}_{-50}$ & $\ddagger3.52^{+0.25}_{-0.25}$ & $0.06^{+0.13}_{-0.13}$ & $0.83^{+0.15}_{-0.15}$ & $9.62^{+4.01}_{-4.01}$ &  &  & $-20.08^{+0.05}_{-0.05}$ & $-13.2^{+0.02}_{-0.01}$ & 517.4 & $0.002^{+0.003}_{-0.001}$ & $3.9^{+2.3}_{-1.0}$ & $1.66^{+0.11}_{-0.14}$ & $41^{+4}_{-4}$ \\
 &  &  & SAND & $1258^{+50}_{-50}$ & $3.98^{+0.25}_{-0.25}$ & $-0.68^{+0.13}_{-0.13}$ &  &  & $0.08^{+0.05}_{-0.05}$ &  & $-19.88^{+0.05}_{-0.05}$ & $-13.6^{+0.02}_{-0.01}$ & 55.8 & $0.013^{+0.021}_{-0.008}$ & $7.1^{+3.6}_{-2.4}$ & $1.37^{+0.12}_{-0.11}$ & $27^{+3}_{-3}$ \\
WISE J073444.02$-$715744.0 & Y0 & (9) & Sonora & $470^{+12}_{-12}$ & $\ddagger3.25^{+0.13}_{-0.13}$ & $-0.33^{+0.25}_{-0.25}$ & $0.35^{+0.12}_{-0.12}$ & $4.6^{+1.5}_{-1.5}$ &  &  & $-19.59^{+0.05}_{-0.05}$ & $-14.84^{+0.02}_{-0.02}$ & 6.0 & $0.016^{+0.012}_{-0.007}$ & $1.2^{+0.4}_{-0.1}$ & $1.29^{+0.05}_{-0.03}$ & $18^{+1}_{-1}$ \\
WISE J082507.35$+$280548.5 & Y0.5 & (16) & Sonora & $357^{+12}_{-12}$ & $\ddagger3.25^{+0.13}_{-0.13}$ & $-0.21^{+0.25}_{-0.25}$ & $0.35^{+0.12}_{-0.12}$ & $5.39^{+1.5}_{-1.5}$ &  &  & $-18.92^{+0.05}_{-0.05}$ & $-14.62^{+0.02}_{-0.02}$ & 7.3 & $\dagger0.041^{+0.042}_{-0.024}$ & $\dagger1.0^{+0.3}_{-0.1}$ & $\dagger1.22^{+0.06}_{-0.04}$ & $8^{+1}_{-1}$ \\
WISEA J085510.74$-$071442.5 & Y2 & (17) & Sonora & $297^{+12}_{-12}$ & $\ddagger3.29^{+0.13}_{-0.13}$ & $-0.85^{+0.25}_{-0.25}$ & $0.65^{+0.13}_{-0.12}$ & $5.74^{+1.52}_{-1.51}$ &  &  & $-18.46^{+0.05}_{-0.05}$ & $-14.57^{+0.02}_{-0.01}$ & 2407.7 & $\dagger0.089^{+0.124}_{-0.054}$ & $\dagger1.0^{+0.3}_{-0.1}$ & $\dagger1.16^{+0.07}_{-0.03}$ & $4^{+1}_{-1}$ \\
ULAS J102940.52$+$093514.6 & T8 & (18) & Sonora & $745^{+25}_{-25}$ & $\ddagger4.08^{+0.16}_{-0.14}$ & $-0.06^{+0.27}_{-0.25}$ & $0.45^{+0.12}_{-0.12}$ & $2.03^{+1.0}_{-1.0}$ &  &  & $-19.8^{+0.05}_{-0.05}$ & $-13.89^{+0.02}_{-0.02}$ & 41.0 & $0.073^{+0.052}_{-0.031}$ & $6.8^{+2.1}_{-1.6}$ & $1.21^{+0.04}_{-0.04}$ & $22^{+1}_{-1}$ \\
 &  &  & LOWZ & $799^{+26}_{-25}$ & $\ddagger3.51^{+0.25}_{-0.25}$ & $0.03^{+0.13}_{-0.13}$ & $0.37^{+0.23}_{-0.23}$ & $1.98^{+1.55}_{-1.51}$ &  &  & $-19.92^{+0.05}_{-0.05}$ & $-13.76^{+0.02}_{-0.01}$ & 85.3 & $0.008^{+0.011}_{-0.005}$ & $2.4^{+1.5}_{-0.9}$ & $1.38^{+0.08}_{-0.07}$ & $29^{+2}_{-2}$ \\
 &  &  & SAND & $830^{+50}_{-50}$ & $5.96^{+0.25}_{-0.25}$ & $0.29^{+0.1}_{-0.1}$ &  &  & $0.02^{+0.05}_{-0.05}$ &  & $-20.01^{+0.05}_{-0.05}$ & $-13.71^{+0.02}_{-0.02}$ & 147.1 & $\dagger10.000^{+0.001}_{-0.001}$ & $\dagger68.1^{+1.0}_{-2.1}$ & $\dagger0.76^{+0.01}_{-0.01}$ & $17^{+1}_{-1}$ \\
CWISEP J104756.81$+$545741.6 & Y0 & (19) & Sonora & $364^{+35}_{-13}$ & $4.92^{+0.13}_{-0.68}$ & $0.78^{+0.15}_{-0.35}$ & $0.37^{+0.12}_{-0.12}$ & $3.71^{+1.05}_{-1.01}$ &  &  & $-19.36^{+0.06}_{-0.14}$ & $-14.98^{+0.02}_{-0.02}$ & 17.9 & $\dagger10.000^{+0.001}_{-5.383}$ & $\dagger25.1^{+11.5}_{-11.0}$ & $\dagger0.90^{+0.09}_{-0.06}$ & $10^{+1}_{-1}$ \\
WISE J120604.38$+$840110.6 & Y0 & (16) & Sonora & $427^{+12}_{-12}$ & $\ddagger3.25^{+0.13}_{-0.13}$ & $-0.11^{+0.25}_{-0.25}$ & $0.34^{+0.12}_{-0.12}$ & $6.0^{+1.5}_{-1.5}$ &  &  & $-19.29^{+0.05}_{-0.05}$ & $-14.7^{+0.02}_{-0.02}$ & 10.6 & $0.022^{+0.019}_{-0.008}$ & $1.1^{+0.4}_{-0.1}$ & $1.26^{+0.03}_{-0.03}$ & $13^{+1}_{-1}$ \\
Ross 458C & T8.5 & (20) & Sonora & $696^{+27}_{-26}$ & $\ddagger3.3^{+0.13}_{-0.13}$ & $-0.6^{+0.25}_{-0.25}$ & $0.26^{+0.12}_{-0.12}$ & $5.85^{+1.52}_{-1.52}$ &  &  & $-20.76^{+0.05}_{-0.06}$ & $-14.82^{+0.01}_{-0.01}$ & 223.1 & $0.006^{+0.003}_{-0.002}$ & $1.6^{+0.4}_{-0.4}$ & $1.39^{+0.04}_{-0.04}$ & $76^{+5}_{-5}$ \\
 &  &  & LOWZ & $800^{+25}_{-25}$ & $\ddagger3.5^{+0.25}_{-0.25}$ & $0.0^{+0.13}_{-0.13}$ & $0.38^{+0.23}_{-0.23}$ & $1.98^{+1.5}_{-1.5}$ &  &  & $-21.05^{+0.05}_{-0.05}$ & $-14.88^{+0.01}_{-0.01}$ & 176.6 & $0.008^{+0.011}_{-0.004}$ & $2.4^{+1.5}_{-0.8}$ & $1.39^{+0.07}_{-0.07}$ & $105^{+8}_{-8}$ \\
 &  &  & SAND & $758^{+50}_{-50}$ & $4.27^{+0.25}_{-0.25}$ & $-0.47^{+0.13}_{-0.13}$ &  &  & $0.08^{+0.05}_{-0.05}$ &  & $-20.89^{+0.05}_{-0.05}$ & $-14.77^{+0.01}_{-0.01}$ & 342.3 & $0.144^{+0.314}_{-0.087}$ & $9.7^{+5.4}_{-3.5}$ & $1.16^{+0.07}_{-0.09}$ & $73^{+7}_{-6}$ \\
SDSSp J134646.45$-$003150.4 & T6 & (21) & Sonora & $946^{+25}_{-25}$ & $\ddagger3.26^{+0.13}_{-0.13}$ & $0.03^{+0.25}_{-0.25}$ & $0.62^{+0.11}_{-0.12}$ & $6.12^{+1.52}_{-1.52}$ &  &  & $-19.61^{+0.05}_{-0.05}$ & $-13.5^{+0.02}_{-0.02}$ & 92.5 & $0.002^{+0.001}_{-0.001}$ & $1.8^{+0.4}_{-0.3}$ & $1.54^{+0.06}_{-0.05}$ & $22^{+2}_{-1}$ \\
 &  &  & LOWZ & $981^{+27}_{-26}$ & $4.51^{+0.26}_{-0.26}$ & $0.5^{+0.13}_{-0.13}$ & $0.53^{+0.23}_{-0.23}$ & $2.76^{+4.01}_{-4.02}$ &  &  & $-19.74^{+0.05}_{-0.05}$ & $-13.33^{+0.01}_{-0.02}$ & 195.6 & $0.215^{+0.254}_{-0.147}$ & $15.7^{+7.9}_{-5.4}$ & $1.11^{+0.09}_{-0.09}$ & $19^{+2}_{-2}$ \\
 &  &  & SAND & $1039^{+51}_{-64}$ & $5.56^{+0.26}_{-0.58}$ & $0.11^{+0.11}_{-0.1}$ &  &  & $-0.01^{+0.03}_{-0.03}$ &  & $-19.78^{+0.1}_{-0.05}$ & $-13.24^{+0.02}_{-0.01}$ & 399.4 & $\dagger10.000^{+0.001}_{-8.316}$ & $\dagger69.1^{+2.1}_{-26.9}$ & $\dagger0.76^{+0.12}_{-0.01}$ & $13^{+2}_{-2}$ \\
WISEPC J140518.40$+$553421.4 & Y0.5 & (16) & Sonora & $391^{+13}_{-12}$ & $\ddagger3.25^{+0.13}_{-0.13}$ & $-0.77^{+0.26}_{-0.25}$ & $0.46^{+0.12}_{-0.12}$ & $5.83^{+1.52}_{-1.51}$ &  &  & $-18.98^{+0.05}_{-0.05}$ & $-14.7^{+0.03}_{-0.03}$ & 2.8 & $\dagger0.029^{+0.030}_{-0.012}$ & $\dagger1.1^{+0.4}_{-0.1}$ & $\dagger1.24^{+0.04}_{-0.04}$ & $9^{+1}_{-1}$ \\
o005\_s41280 & M9-L1 & (2) & Sonora & $2158^{+62}_{-57}$ & $\ddagger3.3^{+0.15}_{-0.13}$ & $-0.39^{+0.27}_{-0.28}$ & $0.71^{+0.22}_{-0.22}$ & $4.07^{+2.87}_{-2.11}$ &  &  & $-23.94^{+0.05}_{-0.06}$ & $-16.82^{+0.04}_{-0.04}$ & 1.8 & $\dagger0.001^{+0.001}_{-0.001}$ & $\dagger8.1^{+31.3}_{-1.9}$ & $\dagger2.24^{+2.51}_{-0.11}$ & $4747^{+2520}_{-2433}$ \\
 &  &  & SAND & $2250^{+52}_{-52}$ & $\ddagger4.14^{+0.27}_{-0.27}$ & $-0.15^{+0.14}_{-0.15}$ &  &  & $-0.03^{+0.03}_{-0.03}$ &  & $-24.01^{+0.05}_{-0.05}$ & $-16.85^{+0.04}_{-0.05}$ & 1.7 & $0.018^{+0.013}_{-0.016}$ & $15.5^{+6.6}_{-4.0}$ & $1.72^{+0.29}_{-0.23}$ & $3916^{+644}_{-603}$ \\
o006\_s00089 & T5-T7 & (2) & Sonora & $926^{+32}_{-33}$ & $4.45^{+0.16}_{-0.18}$ & $0.09^{+0.26}_{-0.26}$ & $0.52^{+0.13}_{-0.13}$ & $3.25^{+1.08}_{-1.11}$ &  &  & $-23.45^{+0.06}_{-0.06}$ & $-17.48^{+0.06}_{-0.08}$ & 1.4 & $0.227^{+0.084}_{-0.142}$ & $13.8^{+4.3}_{-3.4}$ & $1.12^{+0.06}_{-0.06}$ & $1343^{+126}_{-107}$ \\
 &  &  & LOWZ & $953^{+27}_{-25}$ & $5.23^{+0.13}_{-0.13}$ & $0.71^{+0.13}_{-0.13}$ & $0.53^{+0.23}_{-0.23}$ & $2.23^{+4.03}_{-4.02}$ &  &  & $-23.59^{+0.05}_{-0.05}$ & $-17.39^{+0.05}_{-0.05}$ & 1.6 & $2.962^{+2.530}_{-1.181}$ & $46.7^{+9.5}_{-8.1}$ & $0.84^{+0.05}_{-0.05}$ & $1186^{+106}_{-92}$ \\
 &  &  & SAND & $1061^{+53}_{-53}$ & $5.84^{+0.27}_{-0.27}$ & $0.26^{+0.1}_{-0.11}$ &  &  & $0.0^{+0.05}_{-0.05}$ &  & $-23.73^{+0.06}_{-0.06}$ & $-17.23^{+0.04}_{-0.04}$ & 2.2 & $\dagger10.000^{+0.001}_{-0.001}$ & $\dagger71.2^{+1.0}_{-2.1}$ & $\dagger0.76^{+0.01}_{-0.01}$ & $1263^{+90}_{-83}$ \\
o006\_s35616 & L3-L4 & (2) & Sonora & $1819^{+54}_{-56}$ & $\ddagger3.27^{+0.13}_{-0.13}$ & $0.98^{+0.15}_{-0.15}$ & $0.76^{+0.22}_{-0.22}$ & $2.96^{+1.27}_{-1.21}$ &  &  & $-23.29^{+0.06}_{-0.06}$ & $-15.98^{+0.02}_{-0.01}$ & 28.7 & $0.001^{+0.001}_{-0.001}$ & $15.6^{+12.7}_{-11.4}$ & $2.83^{+1.02}_{-0.92}$ & $2830^{+959}_{-956}$ \\
 &  &  & SAND & $1911^{+50}_{-50}$ & $\ddagger3.78^{+0.25}_{-0.25}$ & $-0.46^{+0.13}_{-0.13}$ &  &  & $-0.0^{+0.05}_{-0.05}$ &  & $-23.39^{+0.05}_{-0.05}$ & $-16.32^{+0.02}_{-0.02}$ & 6.5 & $0.003^{+0.004}_{-0.002}$ & $8.2^{+4.5}_{-2.1}$ & $1.77^{+0.23}_{-0.18}$ & $1986^{+264}_{-254}$ \\
CWISEP J144606.62$-$231717.8 & $\geq$Y1 & (19) & Sonora & $351^{+12}_{-12}$ & $\ddagger3.25^{+0.13}_{-0.13}$ & $-0.01^{+0.25}_{-0.25}$ & $0.37^{+0.12}_{-0.12}$ & $5.21^{+1.5}_{-1.5}$ &  &  & $-19.3^{+0.05}_{-0.05}$ & $-15.22^{+0.02}_{-0.03}$ & 4.3 & $\dagger0.044^{+0.044}_{-0.027}$ & $\dagger1.0^{+0.3}_{-0.1}$ & $\dagger1.21^{+0.07}_{-0.03}$ & $12^{+1}_{-1}$ \\
WISE J150115.92$-$400418.4 & T6 & (22) & Sonora & $950^{+25}_{-25}$ & $4.53^{+0.13}_{-0.13}$ & $0.17^{+0.25}_{-0.25}$ & $0.5^{+0.12}_{-0.12}$ & $2.15^{+1.01}_{-1.0}$ &  &  & $-19.82^{+0.05}_{-0.05}$ & $-13.8^{+0.01}_{-0.02}$ & 20.0 & $0.238^{+0.098}_{-0.044}$ & $15.8^{+3.9}_{-3.0}$ & $1.10^{+0.04}_{-0.05}$ & $20^{+2}_{-1}$ \\
 &  &  & LOWZ & $965^{+25}_{-25}$ & $4.5^{+0.25}_{-0.25}$ & $0.39^{+0.13}_{-0.14}$ & $0.53^{+0.23}_{-0.23}$ & $2.72^{+4.02}_{-4.01}$ &  &  & $-19.9^{+0.05}_{-0.05}$ & $-13.64^{+0.02}_{-0.01}$ & 69.7 & $0.220^{+0.229}_{-0.152}$ & $15.1^{+7.7}_{-5.1}$ & $1.11^{+0.09}_{-0.09}$ & $22^{+2}_{-2}$ \\
 &  &  & SAND & $1031^{+50}_{-50}$ & $5.57^{+0.25}_{-0.25}$ & $0.05^{+0.14}_{-0.13}$ &  &  & $0.03^{+0.06}_{-0.06}$ &  & $-19.96^{+0.05}_{-0.05}$ & $-13.43^{+0.01}_{-0.02}$ & 271.7 & $\dagger10.000^{+0.001}_{-6.303}$ & $\dagger69.1^{+2.1}_{-14.0}$ & $\dagger0.76^{+0.06}_{-0.01}$ & $16^{+1}_{-1}$ \\
WISEPA J154151.66$-$225025.2 & Y1 & (16) & Sonora & $373^{+12}_{-12}$ & $\ddagger3.25^{+0.13}_{-0.13}$ & $-0.16^{+0.25}_{-0.25}$ & $0.38^{+0.12}_{-0.12}$ & $5.03^{+1.5}_{-1.5}$ &  &  & $-18.82^{+0.05}_{-0.05}$ & $-14.42^{+0.02}_{-0.02}$ & 7.1 & $\dagger0.035^{+0.036}_{-0.018}$ & $\dagger1.0^{+0.4}_{-0.1}$ & $\dagger1.23^{+0.05}_{-0.03}$ & $7^{+1}_{-1}$ \\
SDSS J162414.37$+$002915.6 & T6 & (7) & Sonora & $988^{+25}_{-25}$ & $4.51^{+0.13}_{-0.13}$ & $-0.01^{+0.25}_{-0.25}$ & $0.43^{+0.12}_{-0.12}$ & $2.05^{+1.0}_{-1.0}$ &  &  & $-19.54^{+0.05}_{-0.05}$ & $-13.32^{+0.02}_{-0.02}$ & 108.7 & $0.209^{+0.070}_{-0.052}$ & $15.5^{+3.5}_{-3.0}$ & $1.11^{+0.05}_{-0.05}$ & $15^{+1}_{-1}$ \\
 &  &  & LOWZ & $969^{+25}_{-25}$ & $4.4^{+0.26}_{-0.26}$ & $0.24^{+0.13}_{-0.13}$ & $0.54^{+0.23}_{-0.23}$ & $3.52^{+4.01}_{-4.01}$ &  &  & $-19.55^{+0.05}_{-0.05}$ & $-13.22^{+0.01}_{-0.02}$ & 228.5 & $0.185^{+0.127}_{-0.142}$ & $12.9^{+6.7}_{-4.7}$ & $1.15^{+0.10}_{-0.09}$ & $16^{+2}_{-2}$ \\
 &  &  & SAND & $1062^{+50}_{-50}$ & $5.68^{+0.25}_{-0.25}$ & $-0.05^{+0.13}_{-0.13}$ &  &  & $0.06^{+0.06}_{-0.05}$ &  & $-19.65^{+0.05}_{-0.05}$ & $-13.09^{+0.02}_{-0.01}$ & 870.7 & $\dagger10.000^{+0.001}_{-3.887}$ & $\dagger70.2^{+1.0}_{-6.1}$ & $\dagger0.76^{+0.02}_{-0.01}$ & $11^{+1}_{-1}$ \\
2MASS J17410280$-$4642218 & L5 & (23) & Sonora* & $1022^{+78}_{-76}$ & $\ddagger3.26^{+0.13}_{-0.13}$ & $0.98^{+0.15}_{-0.16}$ & $0.23^{+0.12}_{-0.12}$ & $8.63^{+0.57}_{-0.79}$ &  &  & $-19.27^{+0.12}_{-0.12}$ & $-12.25^{+0.01}_{-0.02}$ & 14826.4 & $0.002^{+0.001}_{-0.001}$ & $1.9^{+0.5}_{-0.3}$ & $1.59^{+0.06}_{-0.07}$ & $16^{+2}_{-2}$ \\
 &  &  & LOWZ* & $1543^{+71}_{-59}$ & $4.36^{+0.29}_{-0.88}$ & $0.94^{+0.13}_{-0.14}$ & $0.85^{+0.15}_{-0.15}$ & $9.91^{+4.0}_{-4.0}$ &  &  & $-19.97^{+0.06}_{-0.07}$ & $-12.32^{+0.02}_{-0.01}$ & 12310.0 & $0.065^{+0.122}_{-0.059}$ & $14.3^{+14.8}_{-7.3}$ & $1.28^{+0.24}_{-0.21}$ & $28^{+5}_{-5}$ \\
 &  &  & SAND & $1479^{+50}_{-50}$ & $4.0^{+0.25}_{-0.25}$ & $-0.55^{+0.13}_{-0.13}$ &  &  & $0.04^{+0.05}_{-0.05}$ &  & $-19.72^{+0.05}_{-0.05}$ & $-13.14^{+0.02}_{-0.01}$ & 225.5 & $0.010^{+0.016}_{-0.006}$ & $7.9^{+3.7}_{-2.6}$ & $1.43^{+0.14}_{-0.12}$ & $23^{+3}_{-2}$ \\
WISEPA J182831.08$+$265037.8 & Y0 & (6) & Sonora & $375^{+12}_{-12}$ & $\ddagger3.59^{+0.17}_{-0.15}$ & $-0.99^{+0.25}_{-0.25}$ & $0.37^{+0.12}_{-0.12}$ & $7.02^{+0.53}_{-0.53}$ &  &  & $-19.02^{+0.05}_{-0.05}$ & $-14.53^{+0.02}_{-0.01}$ & 1575.3 & $0.127^{+0.093}_{-0.043}$ & $2.2^{+0.8}_{-0.6}$ & $1.19^{+0.01}_{-0.01}$ & $9^{+1}_{-1}$ \\
WISEPA J195905.66$-$333833.7 & T8 & (6) & Sonora & $747^{+25}_{-26}$ & $4.25^{+0.13}_{-0.13}$ & $-0.05^{+0.27}_{-0.25}$ & $0.44^{+0.12}_{-0.12}$ & $2.99^{+1.03}_{-1.15}$ &  &  & $-19.54^{+0.06}_{-0.05}$ & $-13.65^{+0.01}_{-0.02}$ & 66.1 & $0.133^{+0.135}_{-0.052}$ & $9.3^{+2.5}_{-1.9}$ & $1.16^{+0.04}_{-0.04}$ & $15^{+1}_{-1}$ \\
 &  &  & LOWZ & $797^{+25}_{-25}$ & $\ddagger3.51^{+0.25}_{-0.25}$ & $0.0^{+0.13}_{-0.13}$ & $0.37^{+0.23}_{-0.23}$ & $2.5^{+4.02}_{-4.02}$ &  &  & $-19.62^{+0.05}_{-0.05}$ & $-13.52^{+0.02}_{-0.01}$ & 156.1 & $0.009^{+0.011}_{-0.005}$ & $2.5^{+1.4}_{-0.9}$ & $1.38^{+0.08}_{-0.06}$ & $20^{+2}_{-2}$ \\
 &  &  & SAND & $820^{+50}_{-50}$ & $5.92^{+0.25}_{-0.25}$ & $0.29^{+0.1}_{-0.1}$ &  &  & $0.02^{+0.05}_{-0.05}$ &  & $-19.7^{+0.05}_{-0.05}$ & $-13.46^{+0.02}_{-0.02}$ & 246.2 & $\dagger10.000^{+0.001}_{-0.001}$ & $\dagger68.1^{+1.0}_{-3.1}$ & $\dagger0.76^{+0.01}_{-0.01}$ & $12^{+1}_{-1}$ \\
WISEPC J205628.90$+$145953.3 & Y0 & (24) & Sonora & $434^{+12}_{-12}$ & $\ddagger3.25^{+0.13}_{-0.13}$ & $-0.18^{+0.25}_{-0.25}$ & $0.35^{+0.12}_{-0.12}$ & $5.28^{+1.51}_{-1.5}$ &  &  & $-18.86^{+0.05}_{-0.05}$ & $-14.15^{+0.02}_{-0.02}$ & 16.0 & $0.020^{+0.017}_{-0.010}$ & $1.1^{+0.4}_{-0.1}$ & $1.27^{+0.05}_{-0.03}$ & $8^{+1}_{-1}$ \\
WISE J210200.15$-$442919.5 & T9 & (9) & Sonora & $571^{+12}_{-12}$ & $\ddagger3.29^{+0.13}_{-0.13}$ & $-0.37^{+0.25}_{-0.25}$ & $0.34^{+0.12}_{-0.12}$ & $4.03^{+1.5}_{-1.5}$ &  &  & $-19.44^{+0.05}_{-0.05}$ & $-14.35^{+0.02}_{-0.02}$ & 7.2 & $0.010^{+0.005}_{-0.004}$ & $1.4^{+0.4}_{-0.4}$ & $1.34^{+0.03}_{-0.03}$ & $16^{+1}_{-1}$ \\
 &  &  & LOWZ & $548^{+25}_{-25}$ & $5.24^{+0.13}_{-0.13}$ & $-0.05^{+0.13}_{-0.13}$ & $0.83^{+0.15}_{-0.15}$ & $6.66^{+4.01}_{-4.01}$ &  &  & $-19.44^{+0.05}_{-0.05}$ & $-13.98^{+0.02}_{-0.02}$ & 31.8 & $\dagger10.000^{+0.001}_{-0.001}$ & $\dagger40.9^{+5.2}_{-5.2}$ & $\dagger0.83^{+0.02}_{-0.02}$ & $10^{+1}_{-1}$ \\
WISEA J215949.54$-$480855.2 & T9 & (22) & Sonora & $566^{+12}_{-12}$ & $\ddagger3.45^{+0.22}_{-0.14}$ & $-0.53^{+0.25}_{-0.25}$ & $0.23^{+0.12}_{-0.12}$ & $4.31^{+1.5}_{-1.5}$ &  &  & $-19.69^{+0.05}_{-0.05}$ & $-14.64^{+0.03}_{-0.02}$ & 4.1 & $0.017^{+0.014}_{-0.007}$ & $1.9^{+0.7}_{-0.5}$ & $1.30^{+0.03}_{-0.03}$ & $21^{+1}_{-1}$ \\
 &  &  & LOWZ & $539^{+26}_{-26}$ & $5.23^{+0.13}_{-0.13}$ & $-0.25^{+0.13}_{-0.13}$ & $0.7^{+0.16}_{-0.16}$ & $6.1^{+4.01}_{-4.01}$ &  &  & $-19.62^{+0.06}_{-0.05}$ & $-14.2^{+0.01}_{-0.02}$ & 20.5 & $\dagger10.000^{+0.001}_{-0.001}$ & $\dagger40.9^{+5.2}_{-5.5}$ & $\dagger0.83^{+0.03}_{-0.02}$ & $12^{+1}_{-1}$ \\
WISE J220905.73$+$271143.9 & Y0 & (25) & Sonora & $324^{+12}_{-12}$ & $\ddagger3.25^{+0.13}_{-0.13}$ & $-0.19^{+0.25}_{-0.25}$ & $0.36^{+0.12}_{-0.12}$ & $6.84^{+1.5}_{-1.5}$ &  &  & $-18.75^{+0.05}_{-0.05}$ & $-14.76^{+0.02}_{-0.03}$ & 3.3 & $\dagger0.053^{+0.068}_{-0.031}$ & $\dagger1.0^{+0.2}_{-0.1}$ & $\dagger1.20^{+0.06}_{-0.04}$ & $6^{+1}_{-1}$ \\
WISEA J235402.79$+$024014.1 & Y1 & (26) & Sonora & $340^{+13}_{-12}$ & $\ddagger3.4^{+0.13}_{-0.13}$ & $-0.24^{+0.25}_{-0.25}$ & $0.25^{+0.12}_{-0.12}$ & $6.15^{+1.5}_{-1.5}$ &  &  & $-19.04^{+0.05}_{-0.06}$ & $-14.94^{+0.02}_{-0.03}$ & 3.2 & $0.110^{+0.078}_{-0.054}$ & $1.4^{+0.4}_{-0.3}$ & $1.17^{+0.03}_{-0.01}$ & $9^{+1}_{-1}$ \\
\enddata
\tablecomments{The ages, masses, and radii are derived from Monte Carlo sampling of the evolutionary models using the fitted $T_{\mathrm{eff}}$ and $\log g$ values. 
Distances are obtained by jointly sampling $\log(R^{2}/D^{2})$ and the radii, with the median (50th percentile) adopted as the final value, and the 16th and 84th percentiles taken as the lower and upper bounds, respectively. If a dagger symbol ($\dagger$) appears before the age, mass, or radius values, it indicates that more than half of the Monte Carlo samples extend beyond the model grid boundaries. In such cases, the reported values, as well as the corresponding distances, should be treated with caution.
\tablenotetext{a}{If not otherwise specified, the uncertainty in spectral type from this work is assumed to be $\pm1$. See Section \ref{subsec:spttype} for details.}
\tablenotetext{b}{Sources marked with an asterisk (*) next to the model name indicate the corresponding $\chi_r^2$ is about an order of magnitude (i.e., $\sim$10 times) larger than the minimum $\chi_r^2$ obtained among all models for that source. For sources with only a single model fit, we also mark with an asterisk if the fit appears unsatisfactory upon visual inspection. The derived parameters from such models should therefore be interpreted with caution.
\tablenotetext{c}{The $\log g$ values flagged with the symbol ($\ddagger$) are considered anomalous/unphysical fits and should be treated with caution.}}
}
\tablerefs{(0) This work, (1) \citet{2024ApJ...975..162L}, (2) \citet{2025ApJ...980..230T},  (3) \citet{2024ApJ...975...31H}, (4) \citet{2015ApJ...799..203G}, (5) \citet{Mace2013ApJS}, (6) \citet{Kirkpatrick2011ApJS}, (7) \citet{Burgasser2006ApJ}, (8) \citet{2013ApJ...772...79A}, (9) \citet{Kirkpatrick2012ApJ}, (10) \citet{2006MNRAS.373L..60L}, (11) \citet{2004ApJ...610.1045S}, (12) \citet{2014ApJ...782....8I}, (13) \citet{2007MNRAS.381.1077R}, (14) \citet{2015ApJ...814..118B}, (15) \citet{2011ApJ...739...81L}, (16) \citet{Schneider2015ApJ}, (17) \citet{2015ApJ...799...37L}, (18) \citet{Thompson2013PASP}, (19) \citet{Meisner2020ApJ}, (20) \citet{2011MNRAS.414.3590B}, (21) \citet{2000ApJ...531L..57B}, (22) \citet{Tinney2018ApJS}, (23) \citet{2016ApJS..225...10F}, (24) \citet{Cushing2011ApJ}, (25) \citet{Cushing2014AJ}, (26) \citet{Schneider2015ApJ}.}

\end{deluxetable}
\end{longrotatetable}

\subsection{Overview of Derived Physical Parameters} \label{subsec:overview}

\subsubsection{Atmospheric Fits across the L/T Transition}

\begin{figure}[t]
\centering 
\includegraphics[width=0.5\textwidth]{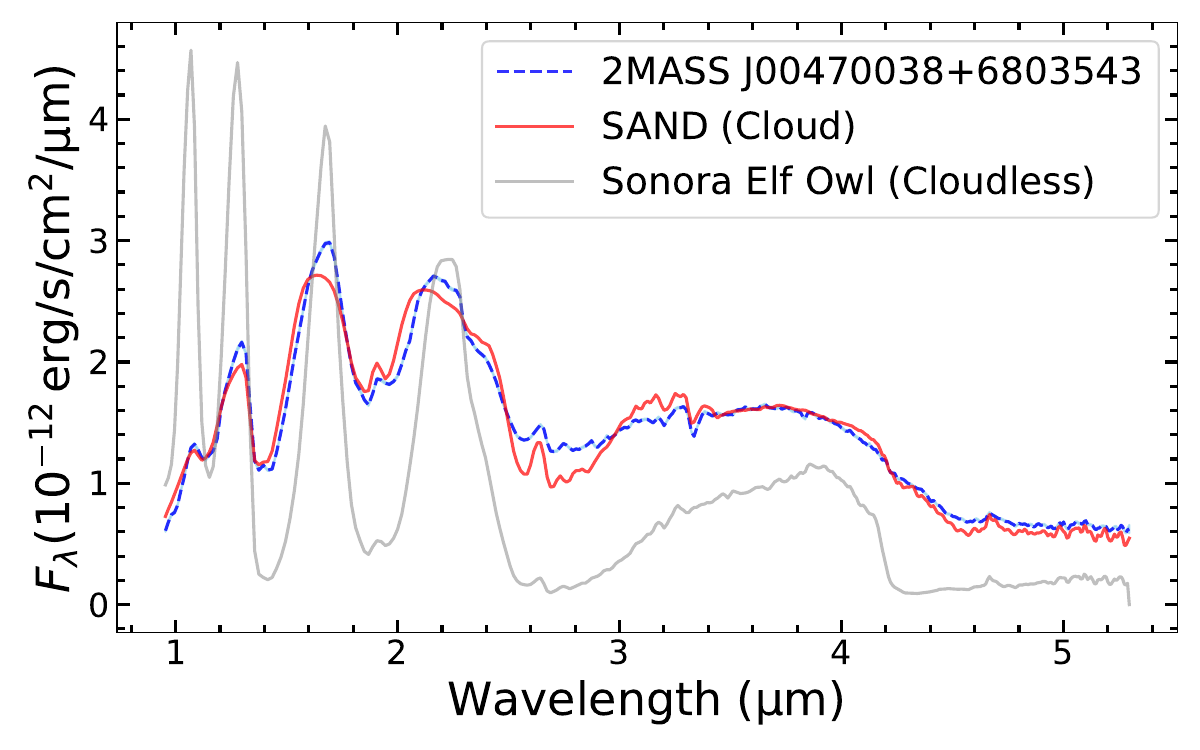}
\caption{Comparison of the observed spectra of 2MASS J00470038$+$6803543 (dashed blue) with the best-fit parameters of cloud-free Sonora Elf Owl model (gray) and the cloudy SAND model (red). \\\label{fig:cloud_spec}} 
\end{figure}

\begin{figure*}[t]
\centering 
\includegraphics[height=1\textheight]{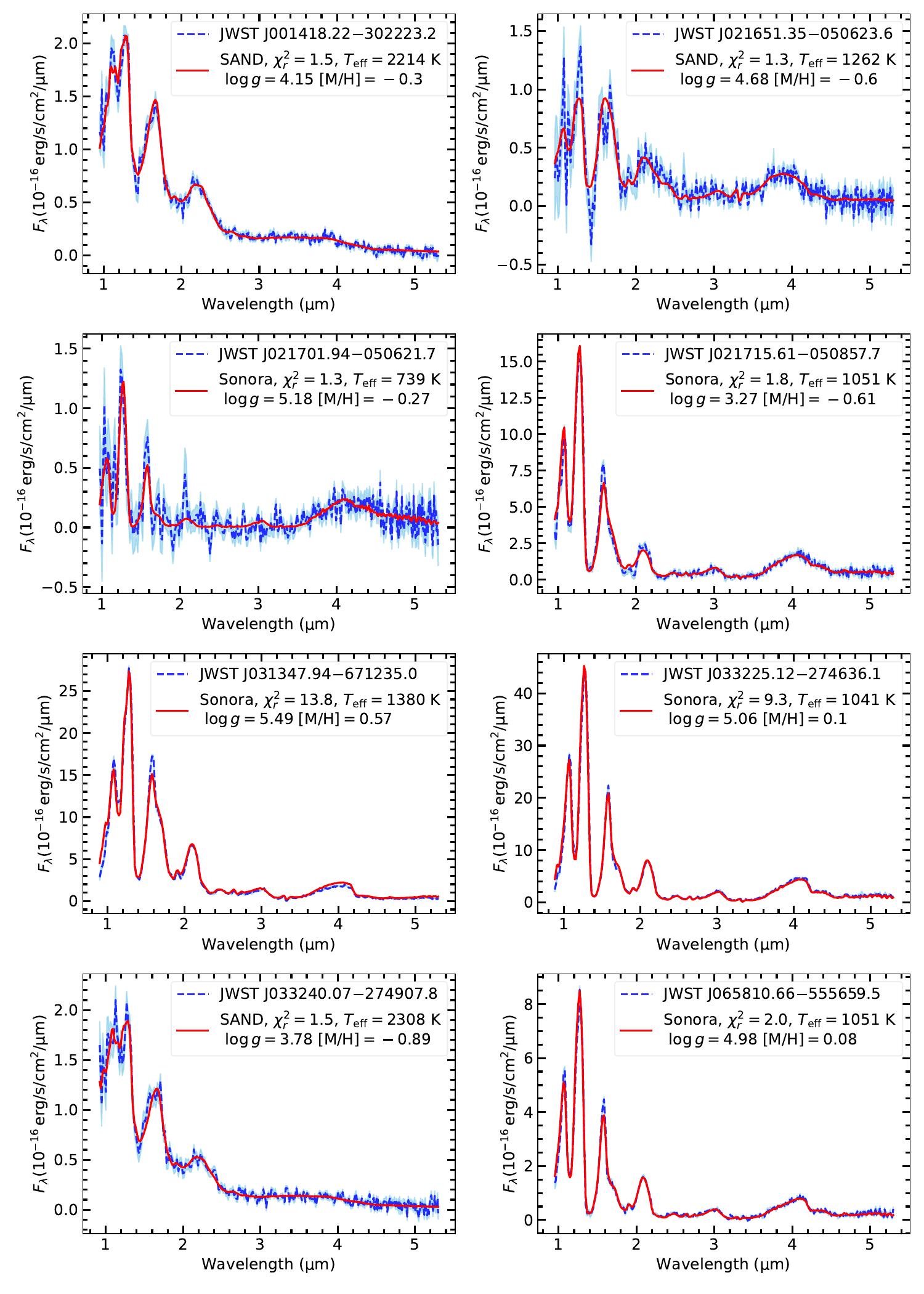}
\caption{JWST/NIRSpec PRISM/CLEAR spectra of the brown dwarf candidates, shown as deep blue lines with light blue shaded regions indicating the uncertainties. The best-fit model spectra are overplotted in red. We show the object names and corresponding best-fit model parameters in the legend.\\\label{fig:bd_spec}} 
\end{figure*}

\begin{figure*}[t]
\ContinuedFloat
\centering 
\includegraphics[height=0.5\textheight]{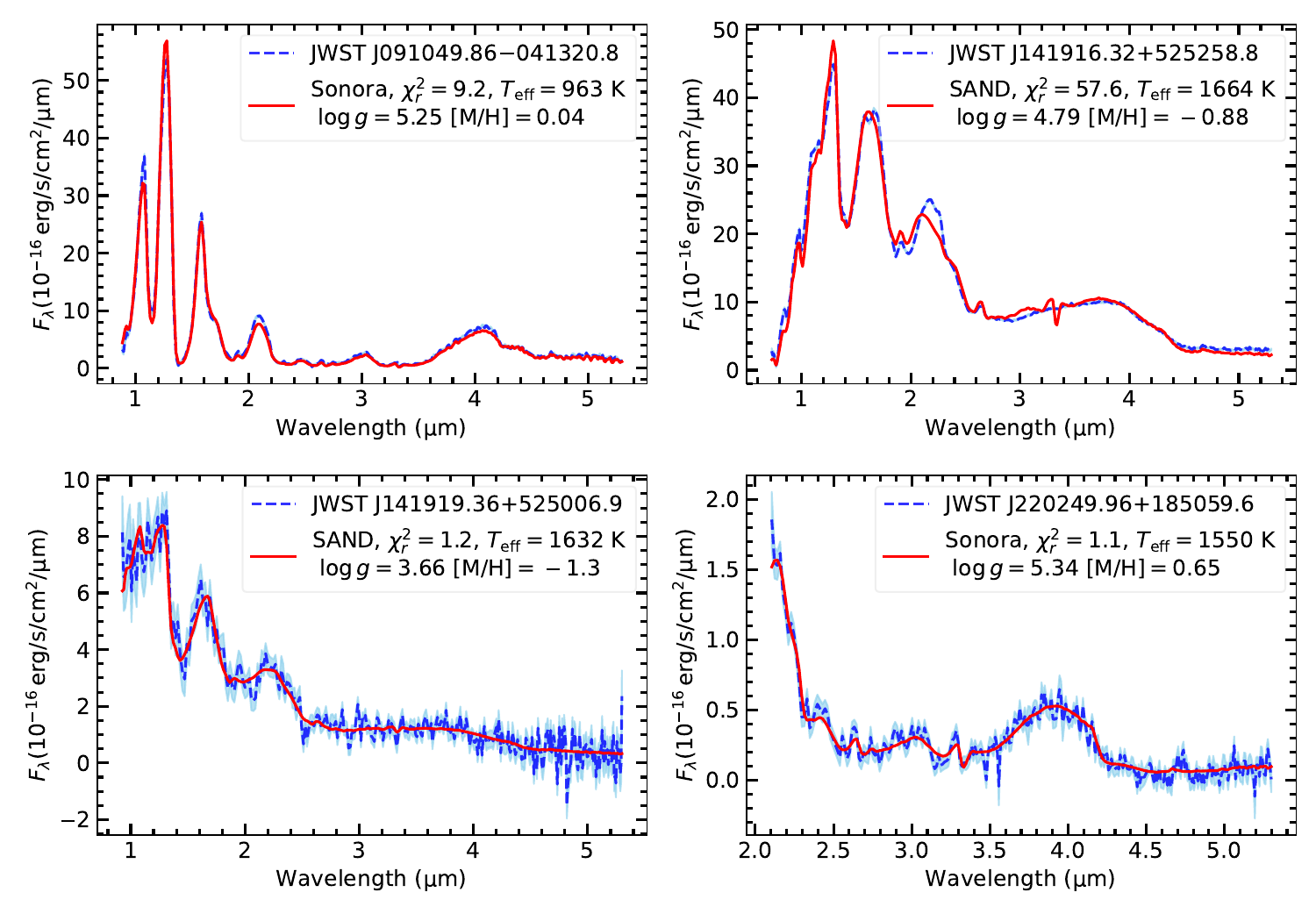}
\caption{(continued)} 
\end{figure*}

We examined the derived physical parameters obtained from all three atmospheric model grids for our entire sample. The $T_\mathrm{eff}$ derived from the different models are generally consistent, with differences typically within 100 K. However, for late-L and early-T dwarfs in the L/T transition, the Sonora Elf Owl and LOWZ (cloud-free) models often yield $T_\mathrm{eff}$ values that differ substantially—by as much as 300-400 K—from those derived using the cloudy model SAND, particularly for objects from PID 3486.

Figure \ref{fig:cloud_spec} shows an example of an L/T transition type brown dwarf in our sample, 2MASS J00470038$+$6803543 (dashed blue), alongside best-fit model spectra from the Sonora Elf Owl (gray) and SAND (red) models. The Sonora Elf Owl model fails entirely to reproduce the dramatic spectral variations induced by atmospheric cloud absorption. As a result, it is unable to fit the observed spectrum of 2MASS J00470038$+$6803543, which exhibits signatures of cloud across the near-infrared wavelength range. In contrast, the best-fit spectrum from the SAND model provides a good match to the observations, capturing both the overall shape and key absorption features.

This failure of cloud-free models to reproduce the observed spectra is consistently observed among all L/T transition-type brown dwarfs in our sample, including brown dwarfs from PID 3486, as well as 2MASS J06462756$+$7935045 and our newly identified object JWST J141916.32$+$525258.8. These objects all have effective temperatures in the range of approximately 1400-1700\,K, yet both the cloud-free Sonora Elf Owl and LOWZ models fail to provide satisfactory fits to their spectra.

For the same set of L/T transition objects, the metallicity estimates from SAND also deviate significantly from those obtained with Sonora Elf Owl and LOWZ. The latter two models tend to predict super-solar metallicities, whereas SAND consistently yields sub-solar values. Similarly, surface gravity inferred from Sonora Elf Owl and LOWZ is systematically lower than that obtained from the SAND model.

\subsubsection{Atmospheric Parameters of the Sonora Elf Owl} \label{subsubsec:sonora}

For the Sonora Elf Owl model version 2, we find that the best-fit $\log g$ derived from the still tend to be systematically lower compared to the other two models \citep{2024ApJ...976...82T}. This discrepancy is not limited to the coldest Y dwarfs, but is also evident among the warmer M/L-type dwarfs.

Moreover, as reported by \citet{2024ApJ...976...82T}, metallicities derived from the Sonora Elf Owl model using NIRSpec-only spectra can differ significantly from those derived using combined NIRSpec+MIRI spectral coverage. We compared our metallicity results from PID 2302 brown dwarfs with those reported in \citet{2024ApJ...976...82T}, and found that such discrepancies mainly occur for late T- and Y-type dwarfs with $T_\mathrm{eff}$ below 600\,K. Therefore, for such sources, the metallicity values derived from the Sonora Elf Owl model should be considered with caution.

\subsubsection{Extinction Parameter}
For brown dwarfs located in young clusters where extinction was included as a free parameter, we compared our derived extinction values with those reported by \citet{2024ApJ...975..162L} for brown dwarfs in PID 1228. We find that our model-fitting results yield $A_{\rm V}$ values that are 2--4 mag higher than those derived by \citet{2024ApJ...975..162L} using standard templates. The systematic offset may arise from several factors. First, standard templates inherently capture the intrinsic spectral energy distributions of young or dusty objects, reducing the required extinction correction, whereas the atmospheric models may underpredict fluxes at shorter wavelengths, causing the fit to compensate by increasing $A_{\rm V}$ \citep{Hurt2024ApJ,2025ApJ...980..230T}. Second, our fitting is performed over the full 0.6-5.3\,$\mu$m NIRSpec range, while the estimates by \citet{2024ApJ...975..162L} are based on a narrower wavelength range $<2.5$\,$\mu$m.

\subsubsection{New Brown Dwarf Candidates}

Overall, the effective temperatures of our new brown dwarf candidates span a wide range from approximately 700\,K—representing the coolest late-type T or potential Y-type candidates—to about 2300\,K, corresponding to early-type L or late-type M dwarfs. Figure~\ref{fig:bd_spec} shows the comparison between the observed spectra and the best-fit model spectra corresponding to the minimum $\chi_r^2$.

For brown dwarf candidates with effective temperatures below 1600\,K, the cloud-free models Sonora Elf Owl and LOWZ yield broadly consistent results for $T_\mathrm{eff}$, $\log g$, [M/H], C/O, and $\log K_{zz}$. In contrast, significant discrepancies arise when comparing the parameters derived from Sonora Elf Owl with those from the cloudy SAND model, as well as between LOWZ and SAND. Their parameter discrepancies primarily arise from whether condensate clouds are included in the atmospheric models, as discussed in the previous section.

\subsection{Spectral Type} \label{subsec:spttype}

We used the \textit{SpeX Prism Library Analysis Toolkit} (SPLAT; \citealt{Burgasser2017ASInC}) to determine the spectral types of each brown dwarf candidate by performing a $\chi^2$ minimization between the observed spectra and a library of standard brown dwarf templates. The standard spectra used for comparison are from \citet{Burgasser2006ApJ}, \citet{Kirkpatrick2010ApJS}, and \citet{Cushing2011ApJ}. 

Following the procedure in \citet{Kirkpatrick2010ApJS}, we carried out the fitting over the 0.95--1.4\,$\mu$m spectral region. The derived spectral types for each candidate are summarized in Table~\ref{tab:allsample}. Unless otherwise noted, the uncertainty on the spectral type is assumed to be $\pm1$ subtype. In cases where the uncertainty is significantly larger, it is explicitly stated. Based on the results, our new brown dwarf candidates consist of eight T-type candidates and four M/L-type candidates.

\subsection{Parameters of the Evolutionary Models}

In this section, we describe the evolutionary models adopted to derive the fundamental parameters of our brown dwarf sample. We use the grids of \citet{Chabrier2023AA}, which span effective temperatures from $\sim$3100\,K for late-M dwarfs down to $\sim$200\,K for the coldest Y dwarfs, making them particularly well suited to our sample. In addition to $T_\mathrm{eff}$, the models also provide predictions for $\log g$, age, mass, radius, and luminosity.

To estimate the physical parameters of individual sources, we interpolated the evolutionary tracks in the two-dimensional space of $T_\mathrm{eff}$ and $\log g$ using the \texttt{LinearNDInterpolator} implementation from \texttt{SciPy}. The input $T_\mathrm{eff}$ and $\log g$ values were adopted from the atmospheric model fits. For each set of $T_\mathrm{eff}$ and $\log g$, we performed 3000 Monte Carlo realizations, from which the 50th percentile was taken as the median value and the 16th and 84th percentiles were adopted as the lower and upper bounds of the uncertainties. The resulting parameters are listed in Table~\ref{tab:allsample}. For input values that fall outside the edges of the evolutionary grids, parameters were derived using the \texttt{NearestNDInterpolator}, i.e., those of the closest grid point. In Table~\ref{tab:allsample}, values preceded by a dagger symbol ($\dagger$) indicate cases where more than half of the Monte Carlo samples ($\geq 1500$ out of 3000) fall outside the model boundaries; in such cases, the \texttt{NearestNDInterpolator} was applied. These results, along with the corresponding distances, should therefore be treated with caution.

In the evolutionary models, age and $\log g$ are strongly correlated, as older brown dwarfs tend to contract and therefore exhibit higher surface gravities. Consequently, a larger $\log g$ generally implies an older age. In our atmospheric model fits, however, we find that the three different model grids often yield noticeably different $\log g$ values for the same source. This discrepancy directly propagates into the evolutionary model inference: since age, mass, and radius are all tightly coupled to $\log g$ within the grids, variations in the fitted surface gravity can lead to substantial differences in the derived physical parameters. For clarity, we have marked with the symbol ($\ddagger$) those cases in Table~\ref{tab:allsample}, where the fitted $\log g$ values appear anomalous or physically implausible, to indicate that these results should be treated with special caution.

\subsection{Distance} \label{subsec:distance}

\begin{figure}[t]
\centering 
\includegraphics[width=0.5\textwidth]{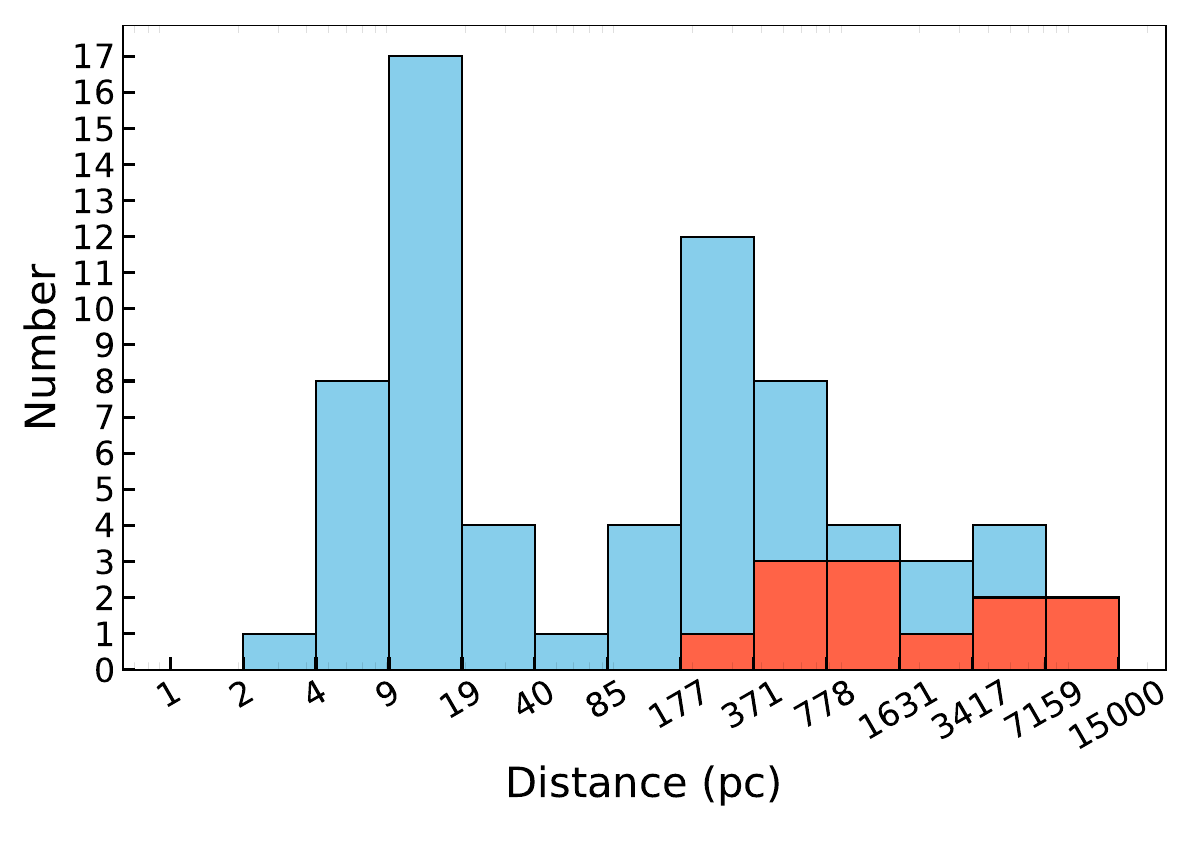}
\caption{The blue histogram shows the distance distribution of all brown dwarfs in our sample, while the red histogram highlights the distribution of twelve new brown dwarf candidates.\\\label{fig:Dis}} 
\end{figure}

The distances of the brown dwarfs were calculated using the radii estimated from the evolutionary tracks together with the $\log(R^2/D^2)$ parameters obtained from the atmospheric model fits. The results are also provided in Table~\ref{tab:allsample}. In addition, we constructed a distance distribution diagram based on the average distance of each source derived from all models, as shown in Figure~\ref{fig:Dis}. This Figure also highlights the distribution of the new brown dwarf candidates. Among the twelve sources, seven are estimated to have distances greater than 1 kpc. Among the cooler T  dwarf candidates---two have distance estimates around or beyond 2\,kpc: JWST J021651.35$-$050623.6, and JWST J220249.96$+$185059.6. 

\begin{figure*}[t]
\centering 
\includegraphics[width=0.5\textwidth]{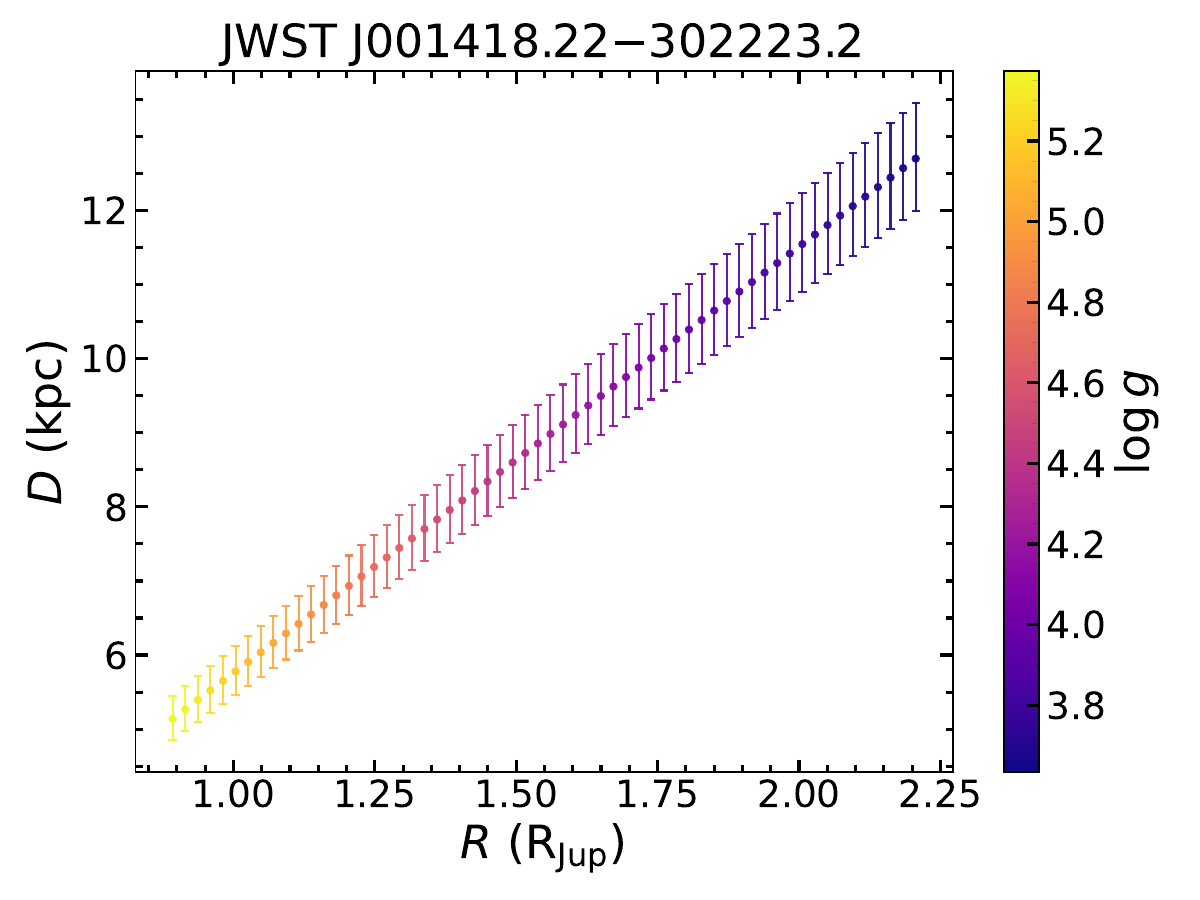}\includegraphics[width=0.5\textwidth]{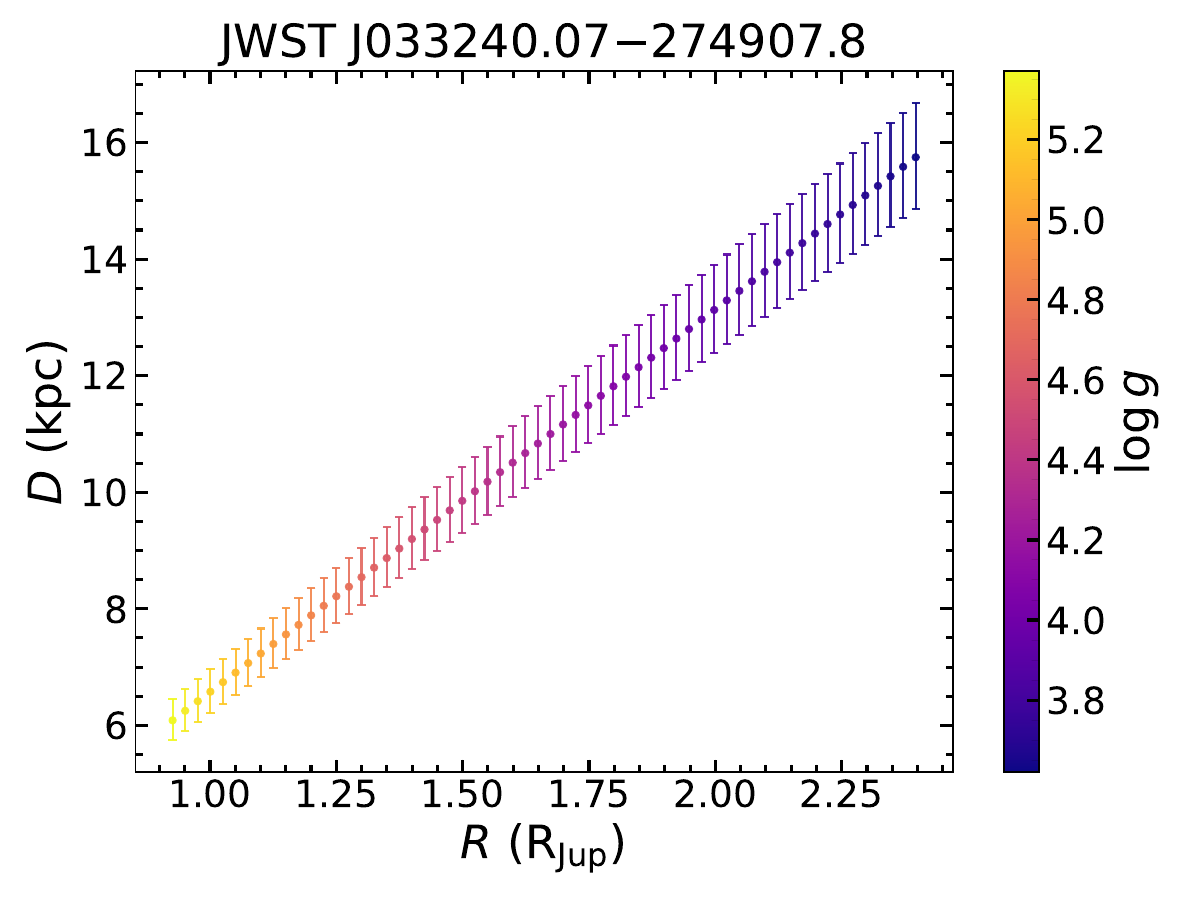}
\caption{Distance estimates using evolutionary tracks for the two most distant brown dwarfs, JWST J001418.22$-$302223.2 (left) and JWST J033240.07$-$274907.8 (right). For each object we fixed $T_\mathrm{eff}$ to 2200\,K and 2300\,K, respectively, and interpolated the evolutionary models over the radius range consistent with $T_\mathrm{eff}\pm50$\,K. Distances $D$ were derived from the measured $\log(R^2/D^2)$ values and the corresponding radii $R$, with vertical error bars reflecting the $\pm0.05$\,dex uncertainty in $\log(R^2/D^2)$. The color coding represents the $\log g$ predicted by the evolutionary models at each radius.\\\label{fig:RD}} 
\end{figure*}

Two of the newly identified brown dwarfs, JWST J001418.22$-$302223.2 and JWST J033240.07$-$274907.8, are inferred from the atmospheric model fits to be located at distances that reach or even exceed 10\,kpc. Both objects, with $T_\mathrm{eff}\sim2200$--2300\,K, represent the most distant spectroscopically confirmed brown dwarf candidates to date. The distances we measured for these two brown dwarfs fall within the detection limits predicted by \citet{2016AJ....151...92R} for L0-type dwarfs with JWST, which extends up to $\sim$16\,kpc. Their metallicities, as derived from both the Sonora Elf Owl and SAND models, are clearly sub-solar, which would strongly suggest membership in the Galactic halo population. However, our age estimates from the same fits indicate unexpectedly young ages, driven by the relatively low $\log g$ values. This is inconsistent with the old ages expected for thick-disk or halo members, which should instead exhibit higher $\log g$ values. A larger $\log g$ would correspond to a smaller radius $R$, and under the same measured $\log(R^2/D^2)$, a smaller radius would yield a shorter distance $D$.

To further assess their distances, we carried out an independent estimate based on evolutionary model radii. For JWST J001418.22$-$302223.2, we adopted $T_\mathrm{eff}=2200$\,K and interpolated the evolutionary models within the temperature range of $2200\pm50$\,K to determine the corresponding range of radii. The associated $\log g$ values were then obtained directly from the evolutionary tracks at those radii. Using the average $\log(R^2/D^2)=-24.81\pm0.05$ obtained from the atmospheric fits, we derived distances in the range 5--13\,kpc, with $D=5.8\pm0.3$\,kpc if $R=1\,R_\mathrm{Jup}$ (Figure~\ref{fig:RD}, left). For JWST J033240.07$-$274907.8, adopting $T_\mathrm{eff}=2300\pm50$\,K and $\log(R^2/D^2)=-29.925\pm0.05$, we obtained distances of 6--16\,kpc, corresponding to $D=6.6\pm0.3$\,kpc if $R=1\,R_\mathrm{Jup}$ (Figure~\ref{fig:RD}, right). Therefore, even under conservative assumptions, both objects are located at distances exceeding 5\,kpc.

\section{Vertical Metallicity Gradient in the Galactic Disk} \label{sec:discuss}

For the metallicity, stars located at large distances---particularly those associated with the Galactic thick disk or halo---are generally expected to exhibit subsolar metallicities \citep[e.g.,][]{2000AJ....119.2843C,2023AJ....166...57M}. This trend also appears to extend to substellar objects: several recently confirmed brown dwarfs at kiloparsec scales have been found to show evidence of metal-poor atmospheres \citep{Burgasser2024ApJ,2024ApJ...975...31H}, supporting the hypothesis that metal-poor brown dwarfs may reside in older Galactic populations. 

Given the relatively large and diverse distance distribution of brown dwarfs in our sample, we are able to examine the possible correlation between metallicity and vertical height above the Galactic plane ($|Z|$), which serves as a proxy for population membership (i.e., thin disk, thick disk, or halo). Here, the vertical height is defined as $Z = d \, \sin b$, where $d$ is the distance and $b$ is the Galactic latitude.

\begin{figure*}[t]
\centering 
\includegraphics[height=0.3\textheight]{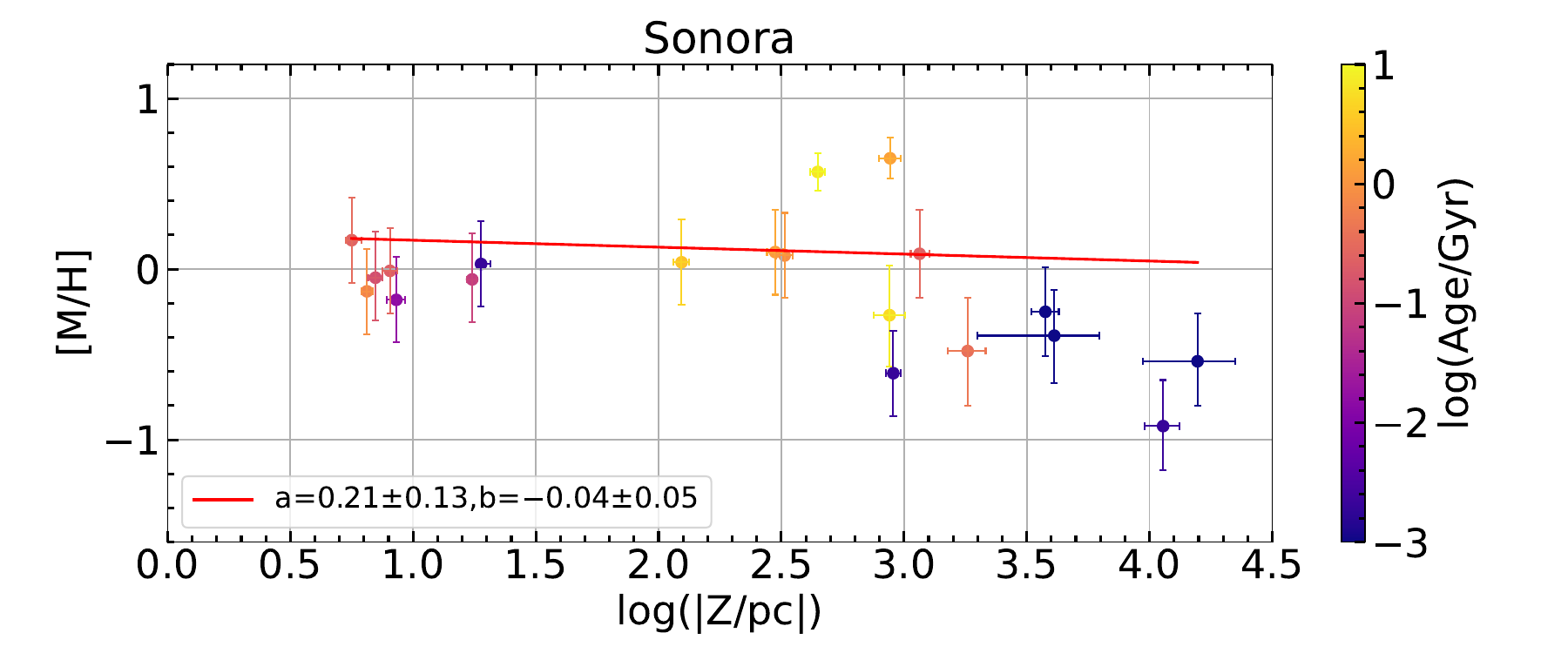}
\includegraphics[height=0.3\textheight]{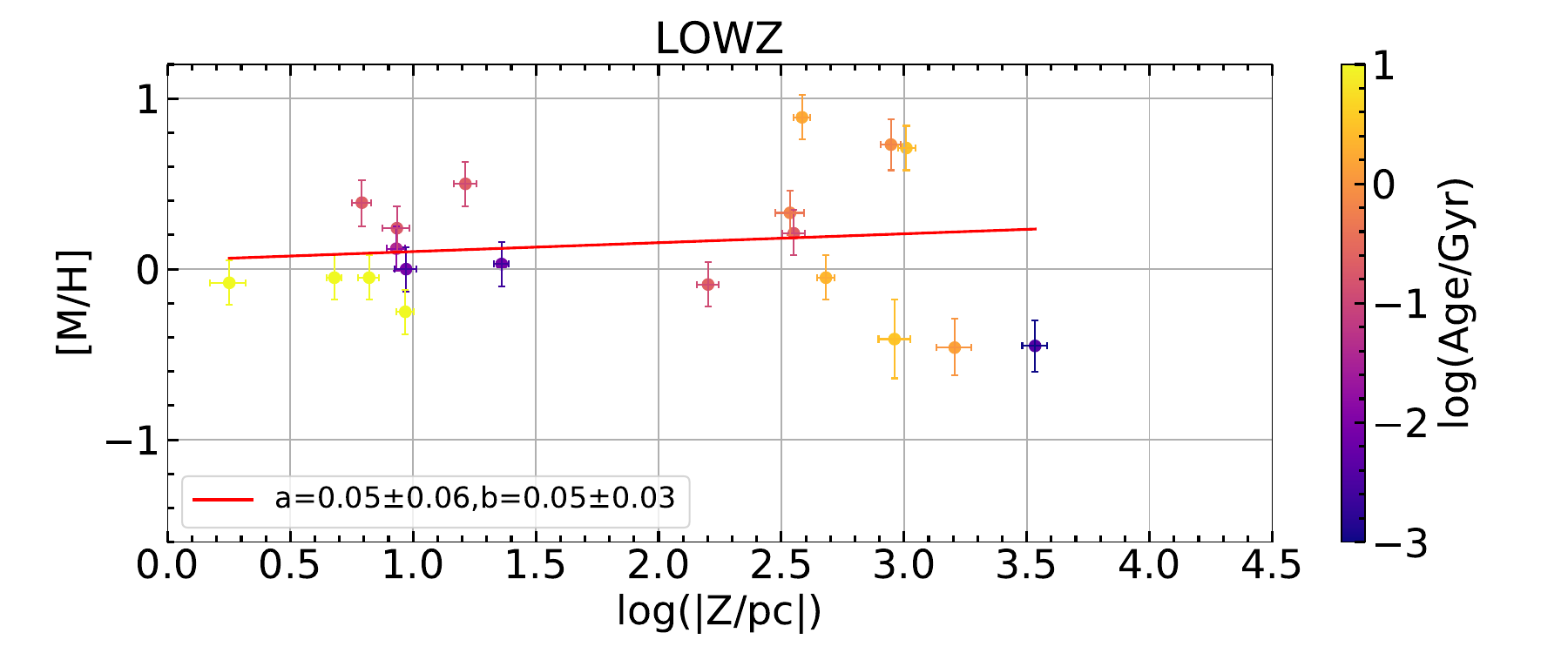}
\includegraphics[height=0.3\textheight]{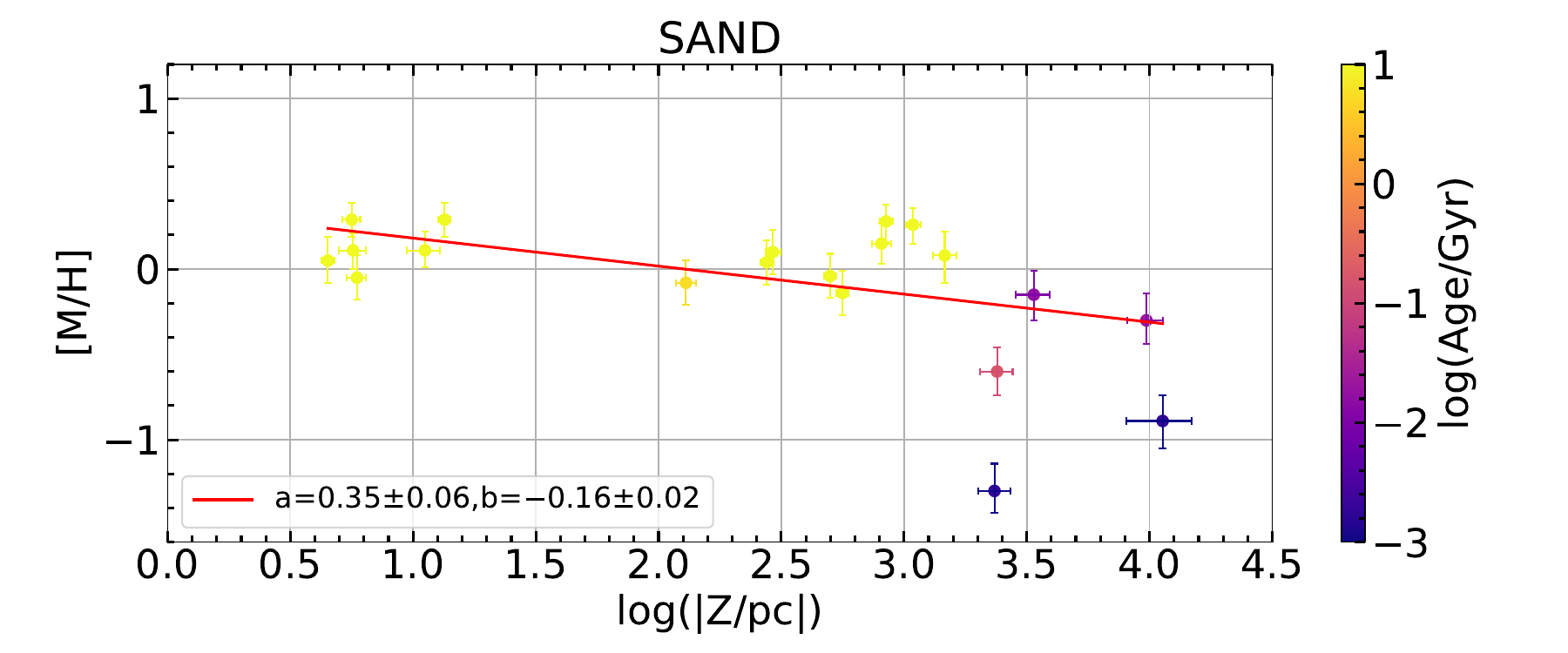}

\caption{The panels show the relationship between metallicity and vertical height $|Z|$, with sources colored by age. The title of each panel specifies the model from which the parameters are derived, while the legend reports the coefficients of the linear fit according to $\mathrm{[M/H]} = a + b \log|Z|$.\\\label{fig:mh_z}} 
\end{figure*}

We constructed the metallicity-$|Z|$ relation using metallicities derived from three atmospheric model grids: Sonora Elf Owl, LOWZ, and SAND. To ensure the reliability of the inferred metallicities, we applied additional selection criteria to the sources included in this analysis. First, we consider only field brown dwarfs and exclude all objects that are located in star clusters. Second, for L/T transition type brown dwarfs (e.g., those from PID 3486), we found substantial discrepancies in the metallicities derived from different model grids, and more importantly, significant mismatches between the observed spectra and the best-fit model spectra, particularly for the Sonora Elf Owl and LOWZ models. As a result, we excluded these objects from this analysis. Third, when using the Sonora Elf Owl model to investigate the metallicity-$|Z|$ relation, we excluded all sources with $T_{\mathrm{eff}} < 600\,\mathrm{K}$ as discussed in Section \ref{subsubsec:sonora}.

After the above selection, we computed the vertical height $|Z|$ for each model using the corresponding distance $d$, and plotted the metallicity against $\log|Z|$ for each case. A simple linear relation,  
\begin{equation}
    [\mathrm{M}/\mathrm{H}] = a + b \, \log|Z| ,
\end{equation}
was fitted to quantify the correlation. The results are summarized in Figure~\ref{fig:mh_z}. To account for the potential impact of stellar age, we used age as the color scale for each point in the plots. The figure shows that those brown dwarfs at larger vertical heights ($\log|Z|\gtrsim3$) are generally inferred to be younger, which is inconsistent with the old ages generally associated with thick-disk and halo populations. As discussed in Section~\ref{subsec:distance}, the atmospheric model fits yield systematically low $\log g$ values for these sources, which in turn drive the inferred ages to be unrealistically young. Therefore, uncertainties or biases in the fitted $\log g$ values may be one of the reasons behind this apparent contradiction.

From the linear fits, both the Sonora and LOWZ models yield slopes around zero ($b \approx 0$). This is because the large metallicity dispersion of brown dwarfs at $\log|Z|\sim2$--4, where both the objects with the highest and lowest metallicities are found. In contrast, most nearby brown dwarfs at $\log|Z|\lesssim2$ have metallicities close to the solar value ($\mathrm{[M/H]}\sim0$). The broad spread at $\log|Z|\sim2$–4 arises from a mixture of populations: metal-poor members of the thick disk or halo, and relatively metal-rich objects whose metallicities may be affected by the model and observed uncertainties. The coexistence of such extremes at large vertical heights weakens any correlation between metallicity and $\log|Z|$.

The SAND model shows a slight trend, with metallicity decreasing as $\log|Z|$ increases. As illustrated in Figure~\ref{fig:mh_z}, the nearby brown dwarfs with $\log|Z| \lesssim 2$ exhibit metallicities close to the solar value, whereas sources at larger vertical heights show relatively lower metallicities, in some cases below $\mathrm{[M/H]} \approx -1$. The linear fit yields a negative slope ($b \approx -0.16\pm0.02$). However, it should be noted that the SAND model imposes an upper metallicity limit of $\mathrm{[M/H]}=0.3$, and two sources around $\log|Z| \approx 3$---JWST J220249.96$+$185059.6 and o006\_s00089---converge to this boundary value. This constraint may explain why this relation using the SAND model shows a trend compared to the Sonora and LOWZ models. Therefore, we conclude that the observed gradient is biased.

To summarize, the analysis of the vertical metallicity gradient based on our brown dwarf sample does not reveal any evident trend. The result of the SAND model is affected by the intrinsic limitation of the SAND grid. And the relations derived from the Sonora Elf Owl and LOWZ models are less apparent, largely due to the presence of these objects with metallicities $\mathrm{[M/H]}>0.5$ at $\log|Z| \sim 3$, which may be affected by the model and observed uncertainties, thereby weakening the overall correlation. The current limitations in sample size, particularly for distant brown dwarfs, and the model-dependent uncertainties in metallicity determinations, further obscure the robustness of the inferred gradient. Future discoveries of more distant brown dwarfs, combined with improved atmospheric models, more precise parameter estimation techniques, and high-resolution spectroscopy, will be essential for reliably characterizing the vertical metallicity gradient in the substellar regime.

\section{Conclusion} \label{sec:conclusion}

In this study, we conducted a systematic analysis of 41283 publicly available JWST NIRSpec PRISM/CLEAR spectra, compiling a comprehensive sample of 68 currently known brown dwarfs with NIRSpec PRISM spectral coverage, including 12 new brown dwarf candidates. To derive the physical parameters of these objects, we performed atmospheric model fitting using a nested sampling Monte Carlo algorithm in combination with three different state-of-the-art brown dwarf model grids: Sonora Elf Owl, LOWZ, and SAND. In addition, we made a preliminary attempt to explore the presence of vertical metallicity gradients in the Galactic brown dwarf population.

Our key findings are as follows:

\begin{enumerate}
    \item We identified 12 new brown dwarf candidates from the NIRSpec PRISM/CLEAR dataset, including 8 T dwarfs and 4 M/L dwarfs. These newly discovered objects are all located at distances greater than 100 parsecs, with 7 of them lying beyond 1 kiloparsecs. This significantly expands the sample of known distant brown dwarfs, especially in the T dwarf regime.
    
    \item Among these, two M/L-type candidates---JWST J001418.22$-$302223.2 and JWST J033240.07$-$274907.8---are located at distances greater than 5\,kpc, making them the most distant brown dwarfs currently known within the Milky Way. Atmospheric model fitting using multiple grids consistently indicates that both objects have sub-solar metallicities, strongly suggesting that they are members of the Galactic halo population. However, the fitted $\log g$ values for both sources are systematically low, leading to unrealistically young age estimates. This is inconsistent with the old ages expected for thick-disk and halo members, and may point to uncertainties or biases in the model $\log g$ determinations.

    \item Our analysis of the vertical metallicity gradient does not reveal an evident correlation between metallicity and Galactic height $|Z|$. The significant dispersion in metallicity across all heights, combined with the limited number of distant brown dwarfs and uncertainties in model-derived parameters, likely masks any underlying gradient.

\end{enumerate}

The low-mass, long-lived substellar objects serve as critical tracers of the chemical enrichment and dynamical history of the Milky Way. However, in order to fully exploit their diagnostic potential, it is essential to determine their fundamental parameters—such as effective temperature, surface gravity, and metallicity—with the highest possible accuracy, within the constraints of current observational capabilities.

Our work opens the door to exploring the vertical metallicity gradient within the brown dwarf population, establishing a foundation for future investigations into the chemical structure and evolutionary history of the substellar component of the Milky Way. Our study emphasizes the importance of both expanding the population of confirmed distant brown dwarfs and refining metallicity determination techniques. These improvements are vital for robustly constraining the vertical metallicity gradient of substellar objects in the Galaxy, thereby enabling a deeper understanding of the structure and evolution of the Galactic disk and halo.

\section*{Acknowledgments} 
We thank the anonymous referee for the constructive comments and suggestions, which helped improve the quality of this paper. This work is supported by the National Natural Science Foundation of China (NSFC) through the projects 12588202, 12373028, 12322306, 12173047, and 12133002. S. W. and X. C. acknowledge support from the Youth Innovation Promotion Association of the CAS with Nos. 2023065 and 2023055. This work is based on observations made with the NASA/ESA/CSA James Webb Space Telescope. The data were obtained from the Mikulski Archive for Space Telescopes (MAST) at the Space Telescope Science Institute.

\section*{Data Availability}
The Sonora Elf Owl models are divided into three submodels based on spectral type: L-type models \citep{mukherjee_2025_15150881}, T-type models \citep{mukherjee_2025_15150874}, Y-type models \citep{mukherjee_2025_15150865}; The LOWZ models \citep{DVNSJRXUO_2021}; The SAND models (version 2) \citep{gerasimov_2024_11582126}; The Sonora Bobcat models \citep{marley_2021_5063476}.

The spectra of the 68 brown dwarfs can be accessed from MAST via \dataset[doi:10.17909/hryb-x508]{https://doi.org/10.17909/hryb-x508}.

\vspace{5mm}


\software{NumPy \citep{numpy}, Scipy \citep{scipy}, Matplotlib \citep{matplotlib}, UltraNest \citep{ultranest}, SPLAT \citep{Burgasser2017ASInC}}

\bibliography{main}{}
\bibliographystyle{aasjournalv7}

\end{document}